\documentclass[aps,prd,superscriptaddress,nofootinbib,twocolumn,floatfix]{revtex4-1}

\usepackage{mathtools}
\usepackage{amsfonts}
\usepackage{amssymb}
\usepackage{mathrsfs}
\usepackage{dsfont}

\usepackage{bbm}
\usepackage{slashed}
\usepackage{amsmath}
\usepackage{bm}
\usepackage{tensor}

\usepackage{graphicx}
\usepackage{color}
\usepackage{colortbl}
\usepackage{array}
\usepackage[dvipsnames]{xcolor}
\usepackage{array,multirow,graphicx}

\usepackage{float}
\usepackage{placeins}
\usepackage{booktabs}
\usepackage{subcaption}
\usepackage{makecell}
\usepackage{tabstackengine}

\usepackage{xspace}
\usepackage{siunitx}
\usepackage{hyperref}
\usepackage{orcidlink}
\usepackage[nameinlink]{cleveref}

\makeatletter
\AddToHook{cmd/appendix/before}{\def\cref@section@alias{appendix}}
\AddToHook{cmd/appendix/before}{\def\cref@subsection@alias{appendix}}
\AddToHook{cmd/appendix/before}{\def\cref@subsubsection@alias{appendix}}
\AddToHook{cmd/appendix/before}{\def\cref@paragraph@alias{appendix}}
\makeatother
\crefname{appendix}{Appendix}{Appendices}
\Crefname{appendix}{Appendix}{Appendices}

\usepackage{bookmark}

\usepackage{xifthen}
\usepackage{xcolor}
\usepackage{enumitem}
\hypersetup{
	colorlinks,
	linkcolor={red!75!black},
	citecolor={blue!75!black},
	urlcolor={blue!75!black}
}
\usepackage{physics}

\usepackage[utf8]{inputenc}

\setkeys{Gin}{width=0.48\textwidth}

\makeatletter
\newsavebox\myboxA
\newsavebox\myboxB
\newlength\mylenA

\newcommand*\xoverline[2][0.75]{%
	\sbox{\myboxA}{$\m@th#2$}%
	\setbox\myboxB\null
	\ht\myboxB=\ht\myboxA%
	\dp\myboxB=\dp\myboxA%
	\wd\myboxB=#1\wd\myboxA
	\sbox\myboxB{$\m@th\overline{\copy\myboxB}$}
	\setlength\mylenA{\the\wd\myboxA}
	\addtolength\mylenA{-\the\wd\myboxB}%
	\ifdim\wd\myboxB<\wd\myboxA%
	\rlap{\hskip 0.5\mylenA\usebox\myboxB}{\usebox\myboxA}%
	\else
	\hskip -0.5\mylenA\rlap{\usebox\myboxA}{\hskip 0.5\mylenA\usebox\myboxB}%
	\fi}
\makeatother

\newcommand{\LEGO}{LEGO\textsuperscript{\textregistered}}

\newcommand{\imag}{\text{i}}

\graphicspath{{./figures/}}

\usepackage{axodraw2}

\usepackage{xifthen}
\usepackage{xcolor}

\newcommand{\gettitle}{Multi-scattering processes and spectral properties of low-energy QCD}

\hypersetup{
	colorlinks,
	linkcolor={red!75!black},
	citecolor={blue!75!black},
	urlcolor={blue!75!black}, 
	pdftitle={\gettitle},
	pdfauthor={Kockler, Pawlowski, Sattler, Schulz, Wessely},
	pdfkeywords={analytic continuation, correlation functions, functional renormalization group, real time, spectral function}, 
	bookmarksopen=true,
	bookmarksopenlevel=2,
	bookmarksnumbered=true
}

\begin{document}

\title{\gettitle}

\author{Konrad~Kockler\,\orcidlink{0009-0001-3897-9707}}
\affiliation{Institut f\"ur Theoretische Physik, Universit\"at Heidelberg, Philosophenweg 16, 69120
	Heidelberg, Germany}

\author{Jan M.~Pawlowski\,\orcidlink{0000-0003-0003-7180}}
\affiliation{Institut f\"ur Theoretische Physik, Universit\"at Heidelberg, Philosophenweg 16, 69120
	Heidelberg, Germany}
\affiliation{ExtreMe Matter Institute EMMI, GSI, Planckstr. 1, D-64291 Darmstadt, Germany}

\author{Franz R.~Sattler \,\orcidlink{0000-0003-1744-9456}}
\affiliation{Fakult{\"a}t f{\"u}r Physik, Universit{\"a}t Bielefeld, 
	Universit{\"a}tsstraße 4, 33615 Bielefeld, Germany}

\author{Ruwen Schulz\,\orcidlink{0009-0007-8789-7234}}
\affiliation{Institut f\"ur Theoretische Physik, Universit\"at Heidelberg, Philosophenweg 16, 69120
	Heidelberg, Germany}
\affiliation{Physikalisches Institut, Universit\"at Heidelberg, Im Neuenheimer Feld 226, 69120
	Heidelberg, Germany}

\author{Jonas~Wessely \,\orcidlink{0009-0006-8700-104X}}
\affiliation{Institut f\"ur Theoretische Physik, Justus-Liebig-Universit\"at Giessen, Heinrich-Buff-Ring 16, 35392 Giessen, Germany}

\begin{abstract}

We compute quark and meson spectral functions in low-energy QCD, including all-order scatterings and decays of pions and the scalar $\sigma$-mode. For low energies the gluons decouple and the dynamics of two-flavour QCD is well-described by a Quark-Meson model. The computations are performed directly in Minkowski space, using the spectral functional renormalisation group. The full mesonic realtime dynamics is captured with the emergent composite approach which is extended here to realtime processes. The inclusion of all-order scatterings and decays is achieved through the self-consistent treatment of the scattering tails and the momentum-dependent resummation of the four-meson vertex via its Bethe-Salpeter equation. We illustrate the importance of higher-order scatterings using the example of the $\pi \to 3\pi$ scattering threshold — the lowest kinematically accessible channel for the pion.

\end{abstract}

\maketitle

\section{Introduction}
\label{sec:Intro}

Over the past decades, non-perturbative functional and lattice QCD approaches have made substantial progress in computing QCD observables in the Euclidean domain, both in the vacuum and in the QCD phase structure; for recent reviews on functional QCD, see \cite{Fischer:2018sdj, Dupuis:2020fhh, Fu:2022gou, Rennecke:2025bcw, Fischer:2026uni, Fischer:2026vkc}. In the past decade, also the description of realtime dynamics in and out of equilibrium with non-perturbative functional methods has made considerable progress.

At sufficiently low energies or momenta, gluons decouple from the off-shell dynamics. In this regime, QCD reduces to an effective theory, whose dynamics is governed by off-shell quantum, thermal, and density fluctuations of quarks and light hadrons, specifically the pions and the scalar $\sigma$-mode in a gluonic background. These dynamical hadronic low-energy degrees of freedom emerge as resonant channels of the full four-quark interaction vertex. The respective dynamics is quantitatively captured within the fRG approach to QCD with emerging composites, developed and used in \cite{Gies:2002hq, Braun:2008pi, Mitter:2014wpa, Braun:2014ata, Rennecke:2015eba, Cyrol:2017ewj, Fu:2019hdw, Fukushima:2021ctq, Ihssen:2024miv, Pawlowski:2025jpg, Fu:2026qnl, Wang:2026xwa}. This approach has been tested and applied to static properties in vacuum QCD, as well as at finite temperature and density. For $\mu_B/T\lesssim 4.5$ the emergent low-energy effective theory contains dynamical quarks, pions, and the scalar $\sigma$-mode as captured by the Quark-Meson (QM) model with or without a Polyakov loop background, as put forward in \cite{Schaefer:2007pw}. For larger chemical potentials, further composites \cite{Braun:2019aow, Haensch:2023sig} (the density mode and diquarks), as well as the moat regime and inhomogeneous condensation patterns become relevant \cite{Fu:2019hdw,Fu:2024rto,Pawlowski:2025jpg}.

One of the advantageous properties of the functional approach, and specifically of the fRG, is its modular structure, with the three main modules being the pure glue sector, the quark-gluon interface, and the pure matter sector. This modularity is called the \LEGO-principle \cite{Ihssen:2024miv}, and is rooted in the one (fRG) or two-loop (DSE) exact diagrammatics. It allows one to study the different subsectors on their own and to directly include the respective results and error analyses into full QCD.

The realtime dynamics of QCD in and out of equilibrium is chiefly relevant for heavy-ion collisions, and requires direct access to timelike correlation functions. In the past decade, the functional realtime approach has made impressive progress, for the respective literature, including scalar- and QM-type models, see the review \cite{Dupuis:2020fhh}, for more recent works see e.g.~\cite{Tripolt:2021jtp,Heller:2021wan, Tan:2021zid, Jung:2021ipc, Roth:2021nrd, Horak:2022aza, Horak:2022myj, Roth:2023wbp, Horak:2023hkp, Roth:2024rbi, Chen:2024lzz, Fu:2024rto, Pawlowski:2024kxc, Topfel:2024iop, Roth:2025hcm, Kockler:2025kdt, Tan:2025bsv, Roth:2026zrs}. 
The present work furthers this progress with two novel ingredients, which are specifically important in the presence of soft modes \cite{Braun:2023qak}.
These near-massless modes are relevant in the vicinity of a sharp chiral crossover, such as that near a possible critical end point of QCD~\cite{Fu:2023lcm}, as well as in the moat regime or a regime with instabilities towards an inhomogeneous phase \cite{Pawlowski:2025jpg}, which may be thermodynamically stable or appear only on certain momentum scales \cite{Pastor-Gutierrez:2026rsy}.
To capture these modes quantitatively, we do the following: \\[-1ex] 

(1) We compute all-order scattering effects in the resonant pion and $\sigma$-mode channels in terms of a full effective potential of these emergent composites.\\[-2ex]

(2) The spectral function is computed fully self-consistently: it is fed back into the flow that generates all orders of decays, which are otherwise missing. Specifically, this concerns, but is not restricted to, the $\pi \to 3\pi$ decay, which has not been covered in previous works in the QM-model. \\[-1ex]

In the present work, we extend the previous realtime analyses in the QM-model with (1,2) and study its behaviour at $\mu_B=0$ and $T=0$. In particular, we investigate the higher order scattering effects, as well as the $\pi \to 3\pi$ decay. The dynamics covered in (1,2) is enhanced for soft pion and $\sigma$-modes, and we expect sizeable effects for $\mu_B/T\gtrsim 4.5$. The investigation of this regime is left to a forthcoming work, where the advances made in the present work are applied within full functional QCD.

\section{Spectral low-energy QCD}
\label{sec:Spectral_Flows}

The gluon decoupling in functional QCD happens at momentum scales $\approx 700$\,MeV. Below this scale, off-shell gluon fluctuations become rapidly irrelevant, see e.g.~\cite{Rennecke:2015, Alkofer:2018guy}. Accordingly, below this scale low-energy QCD is well-described by an effective theory consisting of only quarks and light hadrons with an ultraviolet scale $\Lambda\approx 700$\,MeV in a gluonic background. Note that this also entails that external scales such as temperature $T$ (or rather the lowest Matsubara frequency) and baryon-chemical potential should not exceed this scale by much. Consequently, such a model is only valid at temperatures $T\lesssim 200$\,MeV. In this work, we restrict ourselves to the two-flavour ($N_f=2$) Quark-Meson model in vacuum, i.e. at vanishing temperature and chemical potential. For the sake of simplicity, we drop the strange quark and the gluonic $A_0$-background (Polyakov loop background). The contributions of both are important for quantitative results but do not change the qualitative realtime dynamics. Their inclusion, as well as the extension of the present analysis to the phase structure at finite baryon chemical potential will be done in forthcoming work within functional QCD. 

The classical action of the two-flavour QM-model is a Yukawa action and is given by
\begin{align}\nonumber 
	S_{\text{QM}}[\Phi] = &\,\int_x \Biggl\{\bar{q}\,\left(i\slashed{\partial}+h_\phi \, \phi_a \tau_a \right)\,q \\[1ex]
	&+\frac{1}{2} \left(\partial_\mu \phi_a\right)^2+ m_\phi^2 \rho -c_\sigma\,\sigma \Biggr\}\,, 
\label{eq:Scl_QM}
\end{align}
with the superfield 
\begin{align} 
	\Phi=(q,\bar q,\phi)\,, \qquad q=(u,d)\,,
\label{eq:Phi}
\end{align}
of quarks $q$ and mesonic composites $\phi$. The quark field $q$ in \labelcref{eq:Scl_QM,eq:Phi} combines the light quark fields, and we resort to the strong-isospin symmetric limit with 
\begin{align} 
	m_q=m_u=m_d\,.
\end{align}
The scalar-pseudoscalar mesonic field $\phi$ consists of the $\sigma$-mode and the three pions and $\rho$ is the radial field, 
\begin{align}
	\phi_a= (\sigma,\boldsymbol{\pi})_a\,,\quad \rho = \frac{1}{2} \phi_a\phi_a\,, \quad \tau =\frac12 \left(\mathds{1}\,,\,\imag\,\gamma_5\, \boldsymbol{\sigma}\right)\,,
	\label{eq:phi-rho}
\end{align}
with the O(4) indices $a=0,1,2,3$ and the Pauli matrices $\boldsymbol{\sigma}=(\sigma_1\,,\sigma_2\,,\sigma_3)$. In the strong-isospin symmetric approximation to QCD, all masses of the scalar-pseudoscalar resonances coincide in the symmetric regime, $m^2_{\phi_i}=m^2_\phi$. The linear term $-c_\sigma\,\sigma$ in the last line of \labelcref{eq:Scl_QM} includes explicit chiral symmetry breaking due to the current mass of the light quarks with
\begin{align} 
	c_\sigma = 2 \frac{m_q \,m_\phi^2}{h_\phi} \,.
\label{eq:cSigma}
\end{align} 
The full effective action $\Gamma[\Phi]$ of the QM-model is the quantum analogue of the classical action and is obtained by integrating quantum fluctuations from some initial scale $k=\Lambda$ to $k=0$. The initial effective action $\Gamma_{\Lambda}$ resembles the classical one 
\begin{align} 
	\Gamma_{\Lambda} = S_{\text{QM},\Lambda}[\Phi] + \Delta\Gamma^\textrm{\tiny{RGC}}_{\Lambda}[\Phi;S_{\text{QM}}] \,, 
	\label{eq:ActionUV}
\end{align} 
but also features non-trivial terms that are enforced by RG-consistency (RGC) \cite{Braun:2018svj}, see also \cite{Pawlowski:2005xe, Pawlowski:2015mlf}. These terms do not introduce further parameters but are functions of the parameters in \labelcref{eq:Scl_QM}. RG-consistency also dictates that all parameters in \labelcref{eq:Scl_QM} depend on the initial cutoff scale $\Lambda$ for ensuring $\Lambda$-independence of the physical effective action at vanishing cutoff scale; hence the subscript $\Lambda$. The RG-consistency term contains contributions to the two-point functions with non-trivial momentum dependences that are relevant for the spectral considerations in the current work, see \Cref{sec:GammaLambdaRGC}. Furthermore, it enforces a UV-relevant $\phi^4$-term and a UV-irrelevant four-quark term, 
\begin{align} 
	\Delta\Gamma^\textrm{\tiny{RGC}}_{\Lambda}\propto \int_x \Biggl\{ \frac{\lambda_\phi}{2}\rho^2- \lambda_{q} \left[ \left( \bar q \tau^0 q\right)^2 +\left(\bar q \boldsymbol{\tau} q\right)^2 \right] \Biggr\} \,. 
\label{eq:phi4q}
\end{align}
Specifically, the four-quark term is relevant for the self-consistent analysis in the present work with emergent composites, see~\Cref{sec:EC}. Note also that the RG-consistent initial effective action $\Gamma_{\Lambda}$ is an approximation of the full effective action of QCD after the decoupling of the gluons, which can be read off from the full functional QCD flows, see~\cite{Ihssen:2024miv}. While RG-consistency already implies non-trivial field and momentum dependences, the respective low-energy effective action $\Gamma_{\Lambda}$ carries further field and momentum dependences that cannot be obtained by RG-consistency but are genuine QCD effects. However, they can be shown to be small, see e.g.~\cite{Cyrol:2017ewj, Ihssen:2024miv, Fu:2025hcm}. As we are mainly interested in the qualitative realtime properties, we will not consider these quantitative effects within this work.

\subsection{Dynamics of emergent composites} 
\label{sec:EC}

The pions and the $\sigma$-mode are emergent composite low-energy degrees of freedom in QCD. In contradistinction to other hadronic resonances, they carry the most relevant off-shell dynamics of low-energy QCD, specifically the pions, which are the lightest QCD degrees of freedom. 

In the fRG approach with emergent composites~\cite{Gies:2001nw, Pawlowski:2005xe, Floerchinger:2009uf}, they are dynamically generated. Within its application to QCD, they are defined as resonant interaction channels with the mesonic quantum numbers of the full four-quark scattering vertex. A given tensor channel of this vertex can be reparametrised in terms of a sum of a (mesonic) exchange process in a specific momentum channel and the residual four-quark vertex. For applications to the phase structure of QCD see \cite{Braun:2008pi, Fu:2019hdw, Pawlowski:2025jpg, Fu:2026qnl}.

\begin{figure}[t]
	\centering
	\begin{minipage}{0.99\linewidth}
		\includegraphics[width=\textwidth]{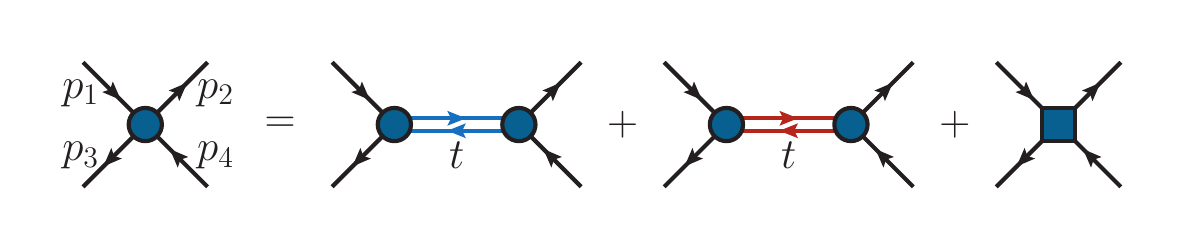}
		\caption{Parametrisation of the four-quark vertex $\Gamma^{(4)}_{qq\bar q \bar q}$ in terms of a $t$-channel exchange of emergent $\pi$ and $\sigma$ composites (scalar-pseudoscalar tensor channel) and the residual vertex $\Gamma^{(4)}_{qq\bar q \bar q,\textrm{res}}$ (blue box vertex). The parametrisation is fixed uniquely by demanding the scalar-pseudoscalar part of the residual vertex to vanish for $s=u=0$ with $s,t,u$ defined in \labelcref{eq:Mandelstam-stu}. See also the discussion in the text around \labelcref{eq:ResonantResidualSplit} to \labelcref{eq:tChannelConstraint}.\hspace*{\fill}}
		\label{fig:bosonisation}
	\end{minipage}
\end{figure}

In the following analysis we follow \cite{Fukushima:2021ctq}. For the pion and $\sigma$-mode, the exchange channel is the $t$-momentum channel of the scalar-pseudoscalar tensor channel.
We depict this in \Cref{fig:bosonisation}. On the left-hand side is the full four-quark vertex, depicted with a blue circle vertex.

For the illustration of this procedure we consider the initial effective action, where we drop the RG-consistent term for the sake of simplicity. This leaves us with the classical action \labelcref{eq:Scl_QM} of the QM-model. This vertex is obtained by taking two quark and two antiquark derivatives of 
\begin{align} 
	\Gamma_{\Lambda}[\bar{q},q]= \Gamma_{\Lambda}\left[\bar{q},q,\phi_\textrm{EoM}[q,\bar q]\right] \,, 
\label{eq:Quark-QM}
\end{align}
where $\phi_\textrm{EoM}[q,\bar q]$ is the solution of the mesonic equations of motion. For $\lambda_\phi=0$, the equations of motions can be solved analytically, and we find 
\begin{align} 
	\phi_\textrm{EoM}[q,\bar q]= -\frac{h_\phi}{m_\phi^2} \bar q \tau q\,.
\label{eq:EoMphi}
\end{align} 
Inserting \labelcref{eq:EoMphi} into \labelcref{eq:Quark-QM} reinstates the explicit quark mass term $ m_q \int_x\bar q q$ and provides the total four-quark interaction term in the initial effective action $\Gamma_{q,\Lambda}$ with
\begin{align} 
	\left.	\Gamma_{\Lambda}[q,\bar q] \right|_{(\bar q q)^2}= -\lambda_{q,\textrm{tot}}\int_x \left[ \left( \bar q \tau^0 q\right)^2 +\left(\bar q \boldsymbol{\tau} q\right)^2\right], 
	\label{eq:Fulllambdaq}
\end{align} 
with 
\begin{align} 
	\lambda_{q,\textrm{tot}}= \frac{h_\phi^2}{2 m_\phi^2}+ \lambda_q\,.
\label{eq:lambdaqtotal}
\end{align} 
Hence, for the initial effective action, the left-hand side of \Cref{fig:bosonisation} carries the scalar-pseudoscalar tensor structure with the coupling strength $\lambda_{q,\textrm{tot}}$. The right-hand side consists of a sum of the resonant exchange in the scalar-pseudoscalar channel and the residual vertex, depicted as a four-quark scattering with a blue-square vertex. 
For our example of the initial effective action, the right-hand side are simply the two parts of the full coupling, multiplied with the scalar-pseudoscalar tensor structure. 

In general, all tensor structures are involved. Projecting on the scalar-pseudoscalar tensor structure with the coupling $\lambda_q$, \Cref{fig:bosonisation} reads in the symmetric phase in the vacuum 
\begin{align}\nonumber 
	\lambda_{q,\textrm{tot}}(p_1,p_2,p_3) = &\,\frac{1}{2} h_\phi(p_1,p_3) G_{\phi}(p_1+p_3) h_\phi(p_2,p_4)\\[1ex] 
	&\,+\lambda_{q,\textrm{res}}(p_1,p_2,p_3) \,, 
\label{eq:ResonantResidualSplit}
\end{align}
where we dropped the $\delta$-functions for momentum conservation for all vertices, e.g.~$\lambda_q(p_1,p_2,p_3,p_4)=\lambda_q(p_1,p_2,p_3)(2\pi)^4\delta(p_1+\cdots +p_4)$ for both $\lambda_{q,\textrm{tot}}$ and $\lambda_{q,\textrm{res}}$. The residual coupling $\lambda_{q,\textrm{res}}(p_1,...,p_3)$ reduces to $\lambda_{q,\Lambda}$ in \labelcref{eq:Scl_QM} at the initial cutoff scale. 

Evidently, such a split is not unique. To begin with, it depends on the chosen full tensor basis of four-quark tensor structures (Fierz ambiguity), as well as the selection of the momentum channel. Here we take the common $t$-channel choice, see \cite{Cyrol:2017ewj}. We define the Mandelstam variables $s,t,u$ with 
\begin{align}
	s &= (p_1 + p_2)^2 = 0\nonumber\\[1ex]
	t &= (p_1 + p_3)^2 = P^2\neq 0\nonumber\\[1ex]
	u &= (p_2 + p_3)^2 = 0\,, 
	\label{eq:Mandelstam-stu}
\end{align}
and fix the above split with the constraint 
\begin{align} 
	\left. 	\lambda_{q,\textrm{res}}(p_1,p_2,p_3)\right|_{u=s=0} \stackrel{!}{=} 0\,.
\label{eq:tChannelConstraint}
\end{align} 
Such a split relates the on-shell Yukawa coupling to the Bethe-Salpeter wave function. For a fully momentum-dependent analysis of the above in vacuum QCD, see \cite{Mitter:2014wpa, Cyrol:2017ewj}, for the general analysis of the fRG approach with emergent composites in QCD, see \cite{Fukushima:2021ctq}. In \cite{Mitter:2014wpa, Cyrol:2017ewj, Fu:2025hcm} it has been shown that the resonant scalar-pseudoscalar channel is absolutely dominant in the vacuum and one may drop the residual vertex, as well as all other tensor structures with no sizeable effects on the results. Moreover, the angular dependence of the Yukawa coupling also has a very mild impact within the flow diagrams due to the angular average of the loop integration. Hence, in the flow diagrams we can approximate 
\begin{align} 
	h_{\phi,k}(p_1,p_3) = h_{\phi,k}(t, \theta_{13} ) \approx h(t) \,,
\label{eq:hphi-Approx}
\end{align}
in a slight abuse of notation. The argument $\theta_{13}$ in \labelcref{eq:hphi-Approx} is the angle between the quark and antiquark momenta $p_1$ and $p_3$, 
\begin{align} 
	\cos\theta_{13} = \frac{p_{1}\cdot p_3}{\| p_1\| \| p_3\|}\,.
	\label{eq:Theta13}
\end{align} 
However, the $t$-channel dependence of the Yukawa coupling can be absorbed completely in the $t$-channel propagator $G_\phi(t)$. In summary, we are led to the approximation
\begin{align} 
	h_{\phi,k}(p_1,p_3) G_{\phi,k}(t) h_{\phi,k}(p_2,p_4) \approx h^2_{\phi,k} G_{\phi,k}(t) \,, 
\label{eq:FinalApproxExchange}
\end{align} 
with $h_{\phi,k}=h_{\phi,k}(t=0)$. This is the emergent composite approach with reduced momentum dependence, which has been developed, tested, and used for the QCD phase structure in \cite{Fu:2019hdw, Pawlowski:2025jpg, Fu:2026qnl}. For more technical details, see~\Cref{sec:setupDynhad}.

\subsection{Spectral functions}
\label{sec:SpecFunc}	

The realtime propagation of (off-shell) light quarks $q=(u,d)$ and mesonic composites $\phi=(\sigma,\boldsymbol{\pi})$ is determined by their realtime propagators $G_\phi$ and $G_q$. Their Källén-Lehmann (KL) spectral representation or lack thereof in QCD and low-energy effective models has been discussed in \cite{Horak:2022aza, Horak:2023hkp, Pawlowski:2024kxc, Kockler:2025kdt}.
In QCD, neither the quark nor the mesonic fields $\phi$ represent asymptotic states: the quark field is not gauge invariant and the field $\phi$ is introduced as an (off-shell) degree of freedom in a scattering vertex. This implies that in QCD their spectral representation may also include a sum over complex conjugate (cc) poles. The present low-energy effective model features a physical cutoff (given by a Landau pole) which also may induce cc poles. In the following, we shall \textit{assume} the absence of cc poles in the spectral representation of both quarks and $\sigma$, $\boldsymbol{\pi}$ and test this assumption with the results. 
 
The propagators are given by the inverse of the full two-point functions, evaluated on the solutions of the equations of motion, $q,\bar q=0$ and $\phi=(\sigma_0,\boldsymbol{\pi}=0)$, 
\begin{align}\nonumber 
	\Gamma_{q\bar q}^{(2)}(p) = &\,Z_q(p)\left[ \imag \slashed{p} +M_q(p)\right] \\[1ex]\nonumber 
	\Gamma_{\pi_i\pi_j}(p) = &\, Z_{\pi}(p) \left(p^2 +m_\pi^2\right)\,\delta_{ij}\\[1ex] 
	\Gamma_{\sigma\sigma}(p) = &\, Z_{\sigma}(p) \left(p^2 +m_\sigma^2\right)\,. 
\label{eq:Gamma2s}
\end{align}
The masses $m_{\pi/\sigma}$ in \labelcref{eq:Gamma2s} are the pole masses of the mesonic composites, defined by 
\begin{align} 
\textrm{Re}\,\Gamma_{\phi_a\phi_a}(p^2=-m^2_{\phi_a})=0\,. 
\label{eq:DefofPoleMass}
\end{align} 
Note that this is just a convenient choice and the split of the mesonic two-point function into $Z_{\phi_a}$ and $m_{\phi_a}$ is not unique. In turn, the split in the quark two-point function \textit{is} unique, as $Z_q$, $M_q$ are the dressing of the Dirac and scalar tensor structures respectively.

The Källén-Lehmann spectral representation for the scalar-pseudoscalar mesonic composites has been discussed in detail in \cite{Horak:2023hkp, Kockler:2025kdt}. It is given by
\begin{align}
	G_{\phi}(p_0,\boldsymbol{p}) = \int_0^\infty\frac{\mathrm{d}\lambda }{\pi}\,\frac{\lambda \ \rho_{\phi}(\lambda,\boldsymbol{p})}{\lambda^2+p_0^2}\,,
\label{eq:Meson-KL} 
\end{align}
with the spatial momentum $\boldsymbol{p}$. The spectral functions $\rho_\phi=(\rho_\sigma, \boldsymbol{\rho}_\pi)$ carry the same momentum dimension as the propagators $G_{\phi}=(G_\sigma,\boldsymbol{G}_\pi)$. For more details we refer to the aforementioned works. 

As discussed in the beginning of this Section, the existence of the Källén-Lehmann spectral representation is only guaranteed for propagators of fields that define asymptotic states. More generally, the KL-representation holds true if the (retarded) propagator only features singularities in the complex half plane with $\textrm{Re}(p_0)< 0$, for a detailed discussion see e.g.~\cite{Cyrol:2018xeq}. However, spectral functions are defined regardless of the existence of a Källén-Lehmann spectral representation as
\begin{align}
	\rho_{\sigma/\pi}(\omega,\boldsymbol{p})=2\,\text{Im}\,G_{\sigma/\pi}\big(p_0\to -i(\omega+i0^+),\boldsymbol{p})\,, 
	\label{eq:rhoMesonFromG}
\end{align}
It is readily proven that \labelcref{eq:Meson-KL} implies \labelcref{eq:rhoMesonFromG}, and the numerical results of the present work are \textit{compatible} with the existence of the KL-representations for the emergent composites $\phi$ in the self-consistent approximation, see \Cref{fig:KL-ExistenceMesons} in \Cref{app:KL-Existence}: There we compare Euclidean propagators obtained from integrating the spectral flow at Euclidean momenta and the spectral one. Both procedures use the analytic properties of the flow differently and the results only agree if the KL-representation holds true. Note that this and similar numerical checks can only show the \textit{compatibility} of the KL-representation within the numerical error bounds.

In the vacuum, we can use Lorentz symmetry to rewrite \labelcref{eq:Meson-KL} more conveniently as 
\begin{align}
	G_{\phi}(p_0,\boldsymbol{p}) = \int_{0_-}^\infty\frac{\mathrm{d}\lambda}{\pi}\, \frac{\lambda\ \rho_{\phi}(\lambda)}{\lambda^2+p^2}\,,\quad \rho_{\phi}(\lambda)=\rho_{\phi}(\lambda,0)\,. 
	\label{eq:Meson-KL-Vacuum} 
\end{align}
At finite temperature and density, Lorentz symmetry is broken by the rest frame and only \labelcref{eq:Meson-KL} holds true. 

For a stable particle, the vacuum spectral function consists of a one-particle pole and a scattering continuum,
\begin{align}
	\rho_\phi(\lambda) = \frac{2\pi}{Z_\phi}\delta(\lambda^2-m_{\phi}^2) + \theta(\lambda^2-m^2_\text{\tiny{scat}})\tilde{\rho}(\lambda)\,, 
	\label{eq:rhopara}
\end{align}
with the pole mass $m_\phi$, see \labelcref{eq:DefofPoleMass}. The spectral normalisation $Z_\phi$ of the pole is obtained from 
\begin{align}
	Z_\phi = \partial_{p^2}\Gamma_\phi^{(2)}(p)\big|_{p^2=-m_\phi^2}\,. 
	\label{eq:ZPhiDef}
\end{align}
Moreover, stable particles that define asymptotic states obey the spectral sum rule 
\begin{align} 
	\int_0^\infty\frac{\mathrm{d}\lambda}{\pi}\,\lambda\,\rho(\lambda,\boldsymbol{p})=1
\label{eq:SpecSumRule} 
\end{align} 
which entails $Z_\phi \geq 1$ and a decay of the Euclidean propagator with $1/p^2$. In general we have 
\begin{align} 
	\int_0^\infty\frac{\mathrm{d}\lambda}{\pi}\,\lambda\,\rho(\lambda,\boldsymbol{p})= \left\{ \begin{array}{rcl} 
		0 & \textrm{for} &\displaystyle\lim_{p_0\to\infty} p_0^2 G_\phi(p_0)=0 \\[2ex]
		1 &\textrm{for} &\displaystyle\lim_{p_0\to\infty} p_0^2 G_\phi(p_0)=1 \\[2ex]
		\infty & \textrm{for} & \displaystyle\lim_{p_0\to\infty} p_0^2 G_\phi(p_0)=\infty 
		\end{array}
	\right. 
\label{eq:SpectralNorm} 
\end{align}
which can be read off from the spectral representation \labelcref{eq:Meson-KL}. In \labelcref{eq:SpectralNorm} we have used that for an ultraviolet limit $Z_\infty/p^2$ with a finite $Z_\infty$ one may always normalise $Z_\infty$ to unity. For a more comprehensive discussion of both ultraviolet and infrared limits see \cite{Cyrol:2018xeq, Bonanno:2021squ}. 
This concludes the brief overview on the spectral properties of the mesonic composites. 

Turning now to the quarks, their spectral representation has been discussed in functional approaches in \cite{Horak:2022aza, Pawlowski:2024kxc}, and we refer for more details to these works. The quark vacuum two-point function in \labelcref{eq:Gamma2s} contains two tensor structures, as does the quark propagator, 
\begin{subequations}
\label{eq:QuarkProp}
\begin{align}
	G_q(p) = i\slashed{p}\, G_q^D(p)+G_q^S(p)\,, 
	\label{eq:QuarkPropDecomposition}
\end{align}
with 
\begin{align}\nonumber 
	G_q^D(p) = &\, \frac{1}{ Z_q(p) }\,\frac{1}{ p^2 +M_q(p)^2 }\,, \\[1ex]
	G_q^S(p) = &\, \frac{1}{ Z_q(p) } \frac{ M_q(p) }{ p^2 +M_q(p)^2 }\,.
\label{eq:GDS}
\end{align}
\end{subequations}
The two scalar coefficient functions have the spectral representations 
\begin{align}\nonumber 
	G_q^D(p)&=\int_0^\infty\frac{\mathrm{d}\lambda}{\pi}\frac{\rho_D(\lambda)}{\lambda^2+p^2}\,,\\[1ex]
	G_q^S(p)&=\int_0^\infty\frac{\mathrm{d}\lambda}{\pi}\frac{\lambda\,\rho_S(\lambda)}{\lambda^2+p^2}\,.
	\label{eq:Quark-KL}
\end{align}
The spectral functions are obtained from the retarded propagator via
\begin{align}\nonumber 
	\rho_D(\omega) & = 2\omega\,\text{Im}\,G_q^D\left(p^0\to -i(\omega+i0^+)\right)\,,\\[1ex]
	\rho_S(\omega) & = 2\,\text{Im}\,G_q^S\left(p^0\to -i(\omega+i0^+)\right)\,.
	\label{eq:rhoQuarkFromG}
\end{align}
With this convention, both spectral functions have the same mass dimension, $[\rho_D]=[\rho_S]=-1$. 
The quark spectral functions are parametrised analogously to that of a stable scalar field \labelcref{eq:rhopara}, with a one-particle pole and a scattering continuum. The (inverse) residue of the single-particle pole is simply given by the wave function evaluated on the pole, and we will write in a slight abuse of notation ${Z_q = Z_q(p^2=-m^2_{q,\text{pole}})}$. 

As in the composite case, the existence of the KL-representation is not guaranteed, as quarks do not define asymptotic states. 
As for the mesonic composites, we find that the assumption of a KL-representation \labelcref{eq:Quark-KL} is compatible with the results in the present self-consistent approximation, see \Cref{fig:KL-ExistenceQuarks} in \Cref{app:KL-Existence}.

\section{Spectral flows in low-energy QCD}
\label{sec:Setup}

We now proceed with a brief description of the computational setup employed in this work. It consists of four parts, whose results feed into each other: \\[-1ex]

(1) The renormalised spectral Callan-Symanzik flow set up in \cite{Fehre:2021eob, Braun:2022mgx} and further developed in \cite{Horak:2023hkp, Kockler:2025kdt, Pawlowski:2025etp}. It is briefly explained below \labelcref{eq:Ren-flow}.
\\[-2ex] 

(2) The spectral Bethe-Salpeter equation for the four-pion vertex as introduced in \cite{Eichmann:2023tjk} in the $\phi^4$-theory, see \Cref{fig:resummation}. \\[-2ex]

(3) The flow of the full mesonic effective potential, which encodes all order scatterings of mesonic composites at vanishing momentum. \\[-2ex] 

(4) The fRG approach with emergent composites \cite{Gies:2001nw, Pawlowski:2005xe, Floerchinger:2009uf}, which takes into account quark and quark-meson rescattering events. \\[-1ex]

The combination of these four parts is conceptually new and enables a fully self-consistent solution of the spectral functions in the Quark-Meson model. This represents an important qualitative step towards a spectral resolution of the QCD phase structure, where soft modes and their amplification of multi-scattering become relevant. In the following, we summarise the differences from the earlier works mentioned above; for more details, see~\Cref{app:AdditionalSetup}.

\subsection{Renormalised Callan-Symanzik flow and shift symmetry}
\label{sec:CSFlow+ShiftSymmetry}

\begin{figure}[t]
	\centering
	\begin{minipage}{\linewidth}
		\centering
		\includegraphics[width=0.9\textwidth]{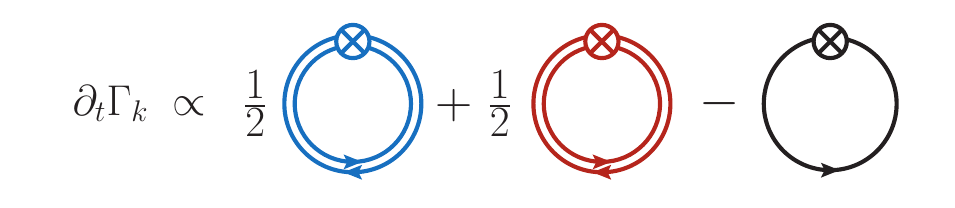}
		\caption{Diagrammatic representation of the flow of the effective action or free energy $\Gamma_k$, omitting the counter term action necessary to render the flow equation with a Callan-Symanzik regulator finite. The notation is explained in~\Cref{fig:notation}. \hspace*{\fill}}
		\label{fig:FlowofGamma}
	\end{minipage}
\end{figure}

\begin{figure}[t]
	\centering
	\begin{minipage}{0.99\linewidth}
		\includegraphics[width=\textwidth]{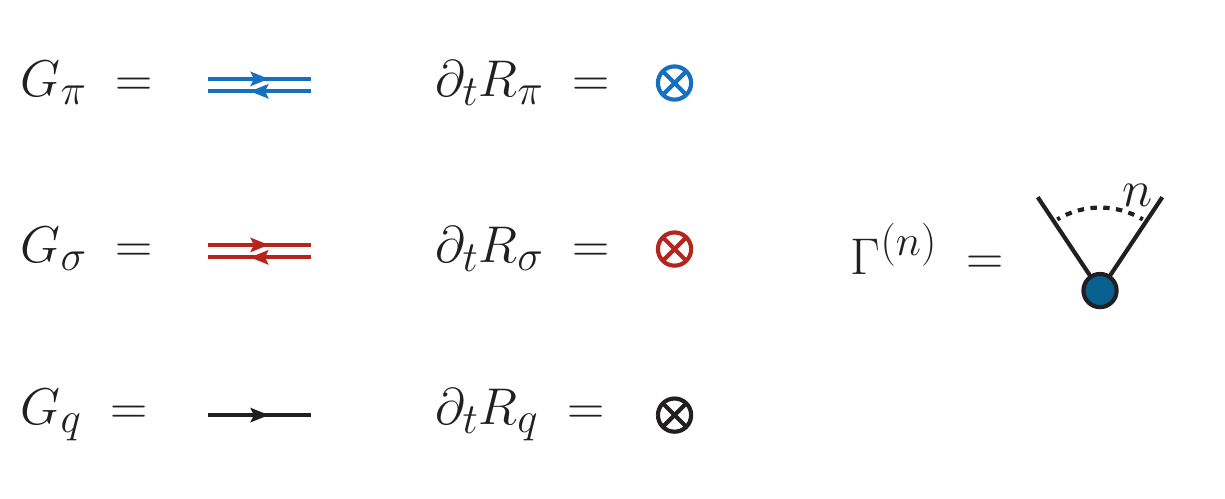}
		\caption{Diagrammatic notation: Full vertices are indicated by blue dots. Full meson propagators are shown as straight double lines: pions (blue), sigma (red). Full quark propagators are shown as straight black lines. Arrows indicate the quark number flow. Regulator insertions $\partial_t R_{\Phi_i}$ carry the corresponding colours. \hspace*{\fill}}
		\label{fig:notation}
	\end{minipage}
\end{figure}

Renormalised flow equations are derived from the standard Wetterich equation with a combination of infrared and ultraviolet regulators, see \cite{Fehre:2021eob, Braun:2022mgx}. It reads 
\begin{subequations}
	\label{eq:Ren-flow} 
\begin{align}
	\partial_t\Gamma_k[\Phi] = \frac{1}{2}\mathrm{Tr}\,G_k[\Phi;R_q,R_\phi]\,\dot{R}_k-\partial_t S_{\text{ct}}\,, 
	\label{eq:Ren-flow1} 
\end{align}
with the full field-dependent propagator 
\begin{align}
	G_k[\Phi;R_q,R_\phi] =\frac{1}{\Gamma_k^{(2)}+R_k}\,,
\label{eq:Gk}
\end{align} 
The flowing counter term $\partial_tS_{\text{ct}}$ in \labelcref{eq:Ren-flow1} is not introduced by hand but follows from a controlled and finite limit of the UV-regularisation. It guarantees the finiteness of the flow equation even for $k\to \infty$, hence describing the flow of renormalised correlation functions. This setup allows us to define a manifestly finite renormalised Callan-Symanzik (CS) flow in the limit 
\begin{align} 
	R_{\phi} \to Z_{\phi} k^2 \qquad R_q\to Z_q k, 
\label{eq:CS-Regs} 
\end{align} 
where $Z_{\phi}$ and $Z_q$ are the wave function of the respective fields on the pole, see \labelcref{eq:ZPhiDef} for the mesonic field and below \labelcref{eq:Quark-KL} for the quarks. 
In contrast to standard Wetterich flows with rapidly decaying regulators, the counter-term action \labelcref{eq:Ren-flow} guarantees both a finite UV limit, and a finite flow. 

The regulator $R_k$ is diagonal in $q-\phi$ space and symplectic in the quark subspace, 
\begin{align} 
	R_k=\textrm{diag}(R_q,R_\phi)\,,\qquad R_q=R^{q\bar q}=-R^{\bar q q}\,,
\end{align} 
for more details see \cite{Ihssen:2024miv} and \Cref{app:AdditionalSetup}. The operator trace on the right-hand side of \labelcref{eq:Ren-flow1} includes mesonic and quark loops, 
\begin{align}\nonumber 
	\frac{1}{2}\mathrm{Tr}\,G_k[\Phi;R_q,R_\phi]\,\dot{R}_k&\\[1ex]
	&\hspace{-3.1cm} = \frac12 \mathrm{Tr}\, G_\phi[\Phi;R_q,R_\phi]\,\dot{R}_\phi-\mathrm{Tr}\,G_q[\Phi;R_q,R_\phi]\,\dot{R}_q\,.
	\label{eq:Ren-flow2} 
\end{align}
\end{subequations}
For a diagrammatic representation of \labelcref{eq:Ren-flow} in the Quark-Meson model, omitting the counter-term action, see~\Cref{fig:FlowofGamma} with the diagrammatic notation \Cref{fig:notation}.

Finally, we discuss the shift-symmetry of the CS-flow which leads to different representations of the same flow, see also \cite{Braun:2022mgx, Topfel:2024iop}. We shall argue that the flow equation \labelcref{eq:Ren-flow} gives access to more fluctuation effects in simpler approximations and hence will be used in the present work. 

For the QM-model with the classical Yukawa action \labelcref{eq:Scl_QM} shift symmetry entails that one may absorb masslike regulators or, more generally, combinations of scalar and pseudoscalar regulators with a shift of the mesonic field, 
\begin{align} 
	\bar{q}\,\left(i\slashed{\partial}+h_\phi \, \phi_a \tau_a + R_q \right)\,q \to \bar{q}\,\left(i\slashed{\partial}+h_\phi \, \phi'_a \tau_a\right)\,q \,,
	\label{eq:AbsorbRk} 
\end{align} 
with the shifted field 
\begin{align} 
	\phi'(x) = \phi(x) + \frac{1}{h_\phi} \,R_q\,, \qquad R_q=\tau_a R_{q,a} \,,
	\label{eq:Shiftphi}
\end{align} 
defined for a momentum-independent regulator. The tensor $\tau$ comprises the scalar-pseudoscalar tensor structures in \labelcref{eq:phi-rho}. Using this shift on the level of the path integral leads us to 
\begin{subequations} 
	\label{eq:ShiftSymfRG}
	\begin{align} 
		\Gamma_{k}[\Phi;R_q,R_\phi]= \Gamma_{k}[\Phi;0,R_\phi] -\frac12 \int_x
		\textrm{tr}\,\left(c_\phi\, \phi\right) \,, 
		\label{eq:ShiftSymAction}
	\end{align} 
	dropping field-independent terms. The trace in the linear term is in Dirac and flavour space and the extension of \labelcref{eq:cSigma}, 
	\begin{align} 
		c_\phi=c_\sigma \tau_0 + 2 \frac{ R_q \,(m_\phi^2+R_\phi)}{h_\phi}\,. 
		\label{eq:cphi}
	\end{align}
	Taking the $t$-derivative of the effective action provides the flow 
	\begin{align}\nonumber 
		\partial_t\Gamma_k =&\, \frac{1}{2}\mathrm{Tr}\,G_\phi[\Phi;0,R_\phi]\dot{R}_\phi\\[1ex] 
		&-\partial_t S_{\text{ct}}+\frac12 \int_x	
		\textrm{tr}\,\left(\partial_t c_\phi\, \phi\right) \,, 
		\label{eq:RenShift-flow} 
	\end{align}
\end{subequations} 
which lacks the quark loop in \labelcref{eq:Ren-flow}. Evidently, the two flows are equivalent as they are derived from the same path integral. In \Cref{app:Shift-DSE+fRG} we show this directly on the level of functional relations by using the mesonic Dyson-Schwinger equation (DSE). The DSEs encode shift symmetry (of the path integral measure) on the level of the effective action. This analysis allows us to single out the more efficient representation if approximations of the effective action are used. 

Seemingly, this analysis advocates the use of \labelcref{eq:RenShift-flow} as it is the simpler equation. This simplicity of the flow comes at the price of accommodating the quark flow on the right-hand side of the flow equation directly via the effective action in terms of its DSE. Consequently, it requires us to include the integral of the quark flow into the effective action instead of having access to the respective fluctuations in form of the flow \labelcref{eq:Ren-flow}. In particular this holds true for momentum dependences. 

In short, \labelcref{eq:Ren-flow} allows for a more efficient use of approximations and speeds up the convergence of approximation schemes. Moreover, for the explicit computation, \labelcref{eq:Ren-flow} can be recast in a more convenient form, 
\begin{subequations} 
	\label{eq:FinalCSFlow} 
	\begin{align}\nonumber 
		\partial_t\Gamma_k[\Phi] + \int_x \dot\phi^a\left(\frac{\delta\Gamma_k[\Phi]}{\delta \phi^a}+c^a_\phi \right) \\[1ex]
		&\hspace{-5.2cm}= \frac{1}{2}\mathrm{Tr} \,G_\phi[\Phi,0,R_\phi] \dot{R}_\phi- \mathrm{Tr} \,G_q[\Phi,0,R_\phi] \dot R_q -\partial_t S_{\text{ct}},\!\! \!
		\label{eq:FinalRen-flow} 
	\end{align}
	with $c_\phi$ in \labelcref{eq:cphi} and hence
	\begin{align} 
		\dot\phi =\frac{1}{h_\phi} \partial_t R_q \,.
		\label{eq:dotphiMain}
	\end{align}
	see also \labelcref{eq:dotphi'} in \Cref{app:Shift-DSE+fRG}. For the CS-flow used in the present work we find 
	\begin{align} 
		\dot R_q = (1- \eta_q)\,Z_q k\,,\quad \dot R_\phi = (2- \eta_\phi)\,Z_\phi k^2\,,
		\label{eq:dotRCS}
	\end{align} 
	and 
	\begin{align} 
		\dot \phi= \frac{1}{h_\phi}(1- \eta_q)\,Z_q k\,.
		\label{eq:dotphiCS}
	\end{align} 
\end{subequations} 
The anomalous dimensions $\eta_q$ and $\eta_\phi$ are defined as the scale derivative of the respective wave functions,
\begin{align}
	\eta_q = -\frac{\partial_t Z_q}{Z_q}\qquad\text{and}\qquad	\eta_\phi = -\frac{\partial_t Z_\phi}{Z_\phi}	
	\label{eq:anomDim}
\end{align}
This concludes our analysis. 

%

\subsection{Approximation of the effective action}
	\label{sec:ApproximationEffAct}

The renormalised CS-flow \labelcref{eq:Ren-flow} is solved in a momentum-dependent approximation that gives access to the spectral properties of the QM-model. The underlying systematic expansion scheme is a combination of the momentum-dependent vertex expansion and the derivative expansion for the mesonic composites: all two-point functions carry their full momentum dependence, and all dressings are introduced with a dependence on the chiral invariant $\rho$ but not its derivatives.
The Yukawa interactions and the self-interactions of the $\sigma$-mode are considered pointlike due to its (infrared) mass being much larger relative to the mass of the pions. 
In turn, the $s$-channel momentum of the four-pion scattering is fully resolved. This self-scattering of the softest mode in QCD feeds back into the flow of the pion two-point function and sources the infrared multi-scattering events. We note in passing that close to the potential critical point in QCD, or more generally the onset of new phases, the $\sigma$-mode takes over the rôle of the softest QCD mode, see \cite{Pawlowski:2025jpg}. Then, one also has to consider the momentum dependence of $\sigma$-vertices to resolve the spectral properties of the matter sector.

Finally, we expand all wave functions about the constant mesonic equation of motion $\rho_0$ and only keep the leading term, $Z_{\Phi_i}(\rho,p)=Z_{\Phi_i}(\rho_0,p)+O(\rho-\rho_0)$. The respective effective action reads 
\begin{subequations} 
	\label{eq:GammaApprox}
\begin{align}\nonumber 
	\Gamma_{k}[\Phi] = &\,\int_p \Biggl\{Z_q(p)\ \bar q(-p)\,\left[\slashed{p} +M_q(p)\right] \, q(p)\\[1ex]\nonumber 
	&\hspace{0.8cm}+\frac{1}{2} Z_\sigma(p)\,\sigma(-p) \left[p^2+m_\sigma^2\right] \,\sigma(p)\\[1ex]\nonumber 
	&\hspace{0.8cm}+\frac{1}{2} Z_\pi(p)\,\boldsymbol{\pi}(-p) \left[p^2+m_\pi^2\right]\,\boldsymbol{\pi}(p) \Biggr\} \\[1ex] \nonumber 
	&\hspace{0.0cm}+\int_x \left( Z_q^{1/2} \bar q \right) h_{\phi,k}\,\left[Z_\phi^{1/2}\left( \phi-\phi_0\right)\right] \, \left( Z_q^{1/2} q\right) \\[1ex]\nonumber 
	& \hspace{0.0cm}+\int_{p} \frac18 \lambda_\phi(p) \left((Z^{1/2}_\pi \boldsymbol{\pi})^2(p)\right)^2 \\[1ex] 
	& \hspace{0.0cm}+
	\int_x V_\textrm{sub} (\rho) - c_\sigma\int_x \sigma\,,
\label{eq:G_QM}
\end{align}
where the $Z(p)=Z(\rho_0,p)$ and $\phi= \phi^a \tau^a$ and all $Z$'s in the Yukawa term are the full momentum-dependent wave functions. Strictly speaking, the EoM $\phi_0$ is also momentum-dependent which is reflected in $M_q(p)$. Since we compute the full momentum dependence of the two-point functions anyway, the momentum-dependence of the solution $\phi_0$ of the EoM is covered. 

The sigma and pion wave functions arise from the full first order term of the derivative expansion with 
\begin{align} 
	\int _x \Biggl\{ Z_\phi(\rho,p) (\partial_\mu \phi^a)^2 + Y_\phi(\rho,p) (\partial_\mu \rho)^2\Biggr\}\,, 
\label{eq:FirstOrderDer}
\end{align} 
evaluated on the solution of the mesonic EoM, 
\begin{align}\nonumber
	Z_\pi(p) =& Z_\phi(\rho_0,p)\,,\\[1ex]
	Z_\sigma(p) =&Z_\phi(\rho_0,p)+\rho_0 Y_\phi(\rho_0,p)\,.
	\label{eq:ZsigmaZpi}
\end{align}
\Cref{eq:ZsigmaZpi} entails that the pion wave function $Z_\pi$ is the dressing $Z_\phi$ of the kinetic term, evaluated on the mesonic EoM, and we shall use this notation. The term $V_\textrm{sub}$ is the full mesonic effective potential with the mass terms and the four-pion term subtracted on the equations of motion, 
\begin{align} \nonumber 
	V_\textrm{sub} (\rho)= &\,V(\rho) -\frac12 Z_\sigma|_{p=0} m^2_\sigma \sigma^2 -\frac12 Z_\pi|_{p=0} m^2_\pi \boldsymbol{\pi}^2 \\[1ex] 
	& -\frac{1}{8} \lambda_\phi \left((Z^{1/2}_\pi \boldsymbol{\pi})^2\right)^2 \,.
\label{eq:Vsub}
\end{align} 
with
\begin{align} 
	Z_\phi|_{p=0}	m^2_\pi =V'(\rho_0)\,,\quad Z_\sigma 	m^2_\sigma = Z_\phi m^2_\pi+2 Z^2_\phi \lambda_\phi \rho_0 \,,
\end{align} 
and 
\begin{align} 
	Z_\phi^2\, \lambda_\phi = V''(\rho_0)\,. 
\end{align} 
\end{subequations} 
All wave functions in \labelcref{eq:Vsub} are evaluated at vanishing momentum. The mesonic mass terms are taken out of $V_\textrm{sub}$ as they are part of the mesonic two-point functions in line two and three of \labelcref{eq:G_QM}. The pointlike four-pion scattering is considered in the $s$-channel approximation in line five of \labelcref{eq:G_QM} with $\lambda_\phi= \lambda_\phi(p=0)$. Note that this split in \labelcref{eq:GammaApprox} is solely introduced to accommodate the four building blocks (1)-(4) discussed at the beginning of \Cref{sec:Setup}. In the following discussions we will always refer to the full effective potential $V$. The flow of the latter is discussed in \Cref{sec:Potential-flow}.

The wave functions are extracted from the momentum dependence of the two-point functions, whose flow consists of several diagrams of the polarisation- and tadpole topology, see \Cref{fig:2pt_flow}. While the former carry genuine momentum dependence already with pointlike interactions, the latter only do so in truncations in which the four-point vertices depend on momentum. Constant contributions are absorbed into the counter terms via the renormalisation condition, see~\Cref{sec:RG-conditions}, thus making these unnecessary to compute in the vacuum.
 
On the contrary, the (tree-level) scattering process $\pi\to 3\pi$, the lowest scattering threshold of the pion, is encoded in the pion tadpole. Dropping this diagram, or equivalently absorbing it into the counter term, removes access to the dominant contribution to this threshold. It survives only via the heavily suppressed intermediate $\pi\to\pi+\sigma\to3\pi$ process. This specific process is part of the pion-sigma polarisation, but only once a full sigma spectral function is included.
Moreover, we have seen in \cite{Kockler:2025kdt} that it is precisely this channel that dominates the spectral function of a scalar particle close to the phase transition, i.e., in the presence of soft modes. Taking this channel into account is thus crucial for describing physics in the QCD phase diagram. Consequently, in order to get access to the dominant part of this lowest scattering threshold of the pion, we use the inhomogeneous Bethe-Salpeter equation to resolve the momentum dependence of the four-pion vertex, see~\Cref{sec:BS-Vertex}. This yields a non-constant pion tadpole, the only tadpole diagram we explicitly compute.

\subsection{RG-consistent initial effective action}
\label{sec:GammaLambdaRGC}

The initial condition for the effective action~\labelcref{eq:GammaApprox} at the initial scale $\Lambda=700$\,MeV is parametrised by~\labelcref{eq:ActionUV}. Its parameters are that of the classical action ~\labelcref{eq:Scl_QM} of the QM-model with pointlike couplings and unit wave functions $Z_{\Phi_i}=1$ for all fields. The RG-consistency term $\Delta\Gamma_{ \textrm{\tiny{RGC}},\Lambda}$ has to be chosen such that the effective action $\Gamma_\textrm{QM} = \Gamma_{\textrm{QM},k=0}$ is $\Lambda$-independent. To begin with, this enforces the inclusion of the UV-relevant $\rho^2$-term as well as the scaling of couplings and wave functions, which scale with their (anomalous) dimension if $\Lambda$ is changed. Moreover, it also enforces higher order terms which are power-counting irrelevant, for more details and examples see \cite{Braun:2018svj}. 

A fully RG-consistent effective action requires both vertex corrections beyond the relevant terms as well as two-point functions with a non-trivial momentum dependence. The latter corrections are important for full spectral RG-consistency: the main reason is the presence of thresholds at large spectral values and specifically at $2m_{q,\Lambda}$ and $m_{\pi,\Lambda}+ m_{q,\Lambda}$ which are not captured by the spectral function of the classical propagator. However, the spectral flow moves them with the cutoff scale to the physical ones. As there is nothing to move due to the lack of RG-consistency, the flow leaves the inverse imprint of them in the full spectral function. This is discussed in detail in \Cref{app:RG-consistency}, see in particular \Cref{fig:rho_Mesons_RGcons} and \Cref{fig:rho_Quark_RGcons}. Accordingly, we cannot drop the non-trivial momentum tail for momenta $p^2\gtrsim \Lambda^2$ in the initial effective action for full RG-consistency. 

These terms could be computed from the requirement 
\begin{align} 
	\Lambda \frac{d\Gamma_{\textrm{\tiny{RGC}},k=0}}{d \Lambda}\stackrel{!}{=}0\,.
\label{eq:RGC}
\end{align} 
However, it is usually more convenient to compute it from integrating the flow up from $k=\Lambda$ to $k=\infty$ in some convergent approximation, typically at one-loop. In the present model this procedure is intricate due to the Landau pole in the meson sector. Moreover, when using a one-loop initial condition, the resulting spectral functions do not obey the spectral sum rule anymore but are normalised according to \labelcref{eq:SpectralNorm} to zero (quarks) or to infinity (mesons). Finally, the present low-energy model can be embedded in full functional QCD and the ultraviolet (spectral) tails of mesonic composite and quarks can be taken from there: \\[-1ex] 

We compute the RG-consistent ultraviolet tail of the quark and meson two-point functions with 
\begin{align} 
	\Delta\Gamma^{(2)}_{\textrm{\tiny{RGC}},\Lambda} \approx \int\limits^\Lambda_\infty \frac{dk}{k} \partial_t \Gamma^{(2)}_\textrm{CS}\bigl[ (h_{\phi,k}, m_{{\phi},k})[\textrm{QCD}]\bigr]\,,
	\label{eq:RGC-1loop}
\end{align} 
with the renormalised Callan-Symanzik flow with the regulators \labelcref{eq:CS-Regs}. For the Yukawa couplings and the meson masses in \labelcref{eq:RGC-1loop} we use the data from \cite{Ihssen:2024miv}. In contrast to the Yukawa coupling of the quark-meson model, which has a classical constant value, the QCD-version decays with $k\to \infty$. This enhances the ultraviolet tail of the Quark-Meson model to that of a QCD-assisted one. The full enhancement would require the use of the QCD running of the Yukawa coupling also for $k\leq \Lambda$ as well as other QCD input. More details concerning \labelcref{eq:RGC-1loop} are provided in \Cref{app:RG-consistency}. 

The QCD-assisted construction used here provides spectral functions whose ultraviolet tails resemble those of QCD, while the infrared part still contains the self-consistent multiscattering dynamics of the low-energy model. Importantly, the decay of the Yukawa coupling in the UV is crucial to circumvent the Landau pole and removes the anomalous running of the wave function renormalisation. With the normalisation prescription employed in the present computation, the resulting spectral functions are normalised to unity. In this context we mention that the UV-flow of the quark is dominated by the quark-gluon interaction, which is missing here. It can be readily included and would change the normalisation of the scalar part of the quark propagator spectral function $\rho_S$ to match the anomalous logarithmic running of the quark mass function in QCD. 

Finally, we drop $\Delta\Gamma_{\textrm{\tiny{RGC}},\Lambda}$ on the right-hand side of the initial flow. Instead, we use the sum of \labelcref{eq:Scl_QM} and \labelcref{eq:phi4q} for the initial flow and add the QCD-assisted initial condition as a global correction. Like the missing quark-gluon diagram, the respective feedback can be readily included, but we leave it to the full functional-QCD analysis.

\subsection{Renormalisation conditions}
\label{sec:RG-conditions}

The inverse propagators of the quarks and mesons in \labelcref{eq:Gamma2s} are computed iteratively, akin to the scheme developed in \cite{Pawlowski:2025etp}. Their CS-flows are made finite by a scale-dependent counter-term action, as derived in \cite{Braun:2022mgx}, whose form is fixed by the renormalisation conditions of the pion and sigma (pole) masses,
\begin{align}\nonumber
	\left[\Gamma_{\pi\pi}^{(2)}+R_{\phi}\right]\big(p^2=-(m_\pi^2 + k^2) \big) &= 0 \\[2ex]
	\text{Re}\left[\Gamma_{\sigma\sigma}^{(2)}+R_{\phi}\right]\big(p^2=-(m_\sigma^2 + k^2) \big) &= 0\,.
	\label{eq:RenormCondMesons}
\end{align}
At $k=0$, these reduce to the physical conditions for the pion and sigma (pole) masses, i.e., $m_{\pi} = 138$\,MeV and $m_{\sigma} = 500$\,MeV. Note that the latter is, at vanishing cutoff, not an asymptotic bound state but a broad resonance, hence the condition is only imposed on the real part of the sigma two-point function. The respective zero crossing of the real part is sometimes referred to as the "Breit-Wigner" mass. For the quarks, we fix the $k$-dependent constituent mass, i.e., the mass function at zero momentum,
\begin{align}
	M_{q,k}(p=0) = M_q(p=0) + k\,,
\end{align}
which coincides at the physical point with the constituent quark mass of full QCD of roughly $M_q(p=0)=350$\,MeV, see e.g.~\cite{Mitter:2014wpa, Cyrol:2017ewj, Ihssen:2024miv, Fu:2025hcm, Ferreira:2026gbe}. 

In contrast to corresponding computation with Dyson-Schwinger equations, or perturbation theory, the diagrams that comprise the flow of the wave-functions are finite for both quarks and mesons. This is due to the fact that the insertion of the CS-regulator lowers the degree of divergence of the diagrams by two. Hence, the integrated flows of the wave functions decay for large momenta. This allows us to let the wave function renormalisation flow freely without imposing any explicit conditions. The renormalisation conditions are implicitly set at the initial scale $\Lambda$ and chosen such that $Z_{\Phi_i}(p^2\to\infty) \to 1$ for all cutoff scales.

As is usual for renormalised CS-flows, any constant contribution of the diagrams to the flows of two-point functions is absorbed in the respective counter terms. In particular, this removes the contribution of constant vertices in the tadpole diagrams.

\subsection{Four-pion \texorpdfstring{$s$}{s}-channel vertex}
\label{sec:BS-Vertex}

\begin{figure}[t]
	\centering
	\begin{minipage}{\linewidth}
		\centering
		\includegraphics[width=0.9\textwidth]{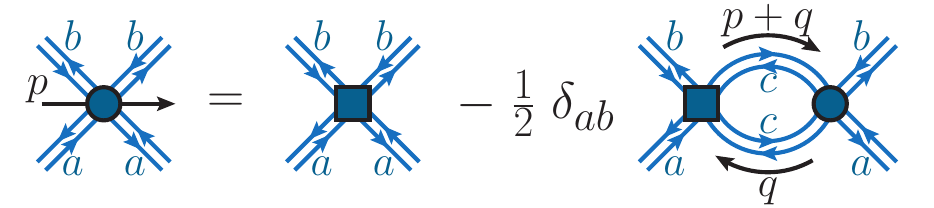}
		\caption{Bubble resummation with the dominant term in the $s$-channel of the $1/N$-expansion to capture the momentum dependence of the four-pion vertex. The blue dot represents the full $s$-channel momentum-dependent vertex, while the blue square represents the RG-improved Bethe-Salpeter scattering kernel. Pion indices $a,b,c$ are indicated in blue. See~\Cref{fig:notation} for the notation.\hspace*{\fill}}
		\label{fig:resummation}
	\end{minipage}
\end{figure}

We follow the technique used in \cite{Horak:2020eng, Braun:2022mgx, Horak:2023hkp, Eichmann:2023tjk, Kockler:2025kdt}, and use the inhomogeneous Bethe-Salpeter equation with an RG-improved scattering kernel, depicted in \Cref{fig:resummation}. Crucially, we identify the latter with the four-pion vertex captured by the effective potential $V$. 
The usage of $V''(\rho)$ as the scattering kernel in \Cref{fig:resummation} ensures that the momentum-dependent vertex coincides at vanishing momentum with the one obtained from the effective potential. 
The full derivation is performed in~\Cref{sec:AdditionalFourPointFunction}, and here we only quote the result, 
\begin{align}
	\Gamma^{(4\pi)}(p) = \frac{V''}{1+\frac{3}{2}V''D_{\text{fish}}(p)}\,.
	\label{eq:pion_four_point_fct}
\end{align}
Our approximation is equivalent to an RG-improved resummation of the leading order in a $1/N$ expansion of the 2PI effective action. In this case, only the tensor structure with four identical pion indices has a non-trivial momentum dependence. $D_{\text{fish}}$ is the fish diagram on the right-hand side of the resummation, renormalised at zero momentum by a simple subtraction. The four-pion vertex defined in that way admits a spectral representation analogous to that of the scalar propagator and is presented in~\cref{sec:AdditionalFourPointFunction}.

\subsection{Mesonic potential}
\label{sec:Potential-flow}

\begin{figure}[t]
	\centering
	\begin{minipage}{0.99\linewidth}
		\includegraphics[width=\textwidth]{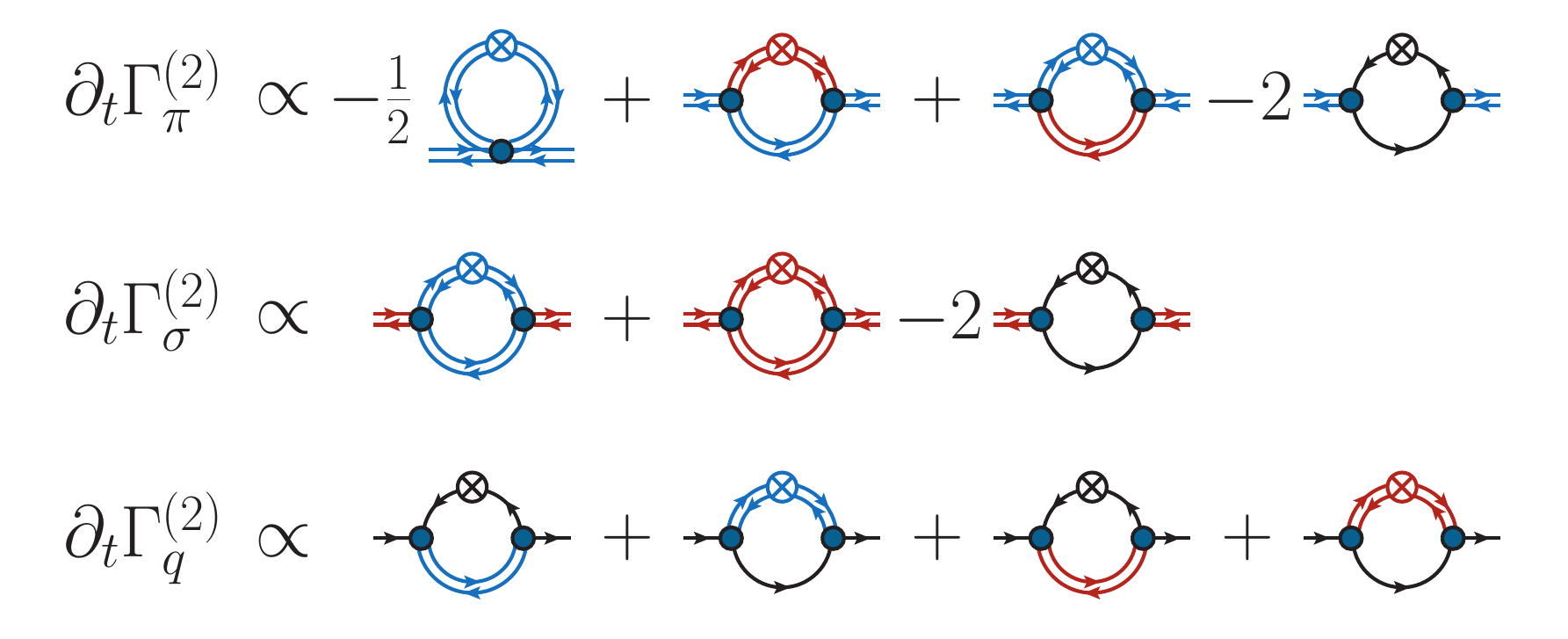}
		\caption{Momentum-dependent diagrams making up the flows of the inverse pion, sigma, and quark propagators in the used truncation. The notation is explained in~\Cref{fig:notation}. Notably, the pions are the only particles for which a momentum tadpole is taken into account. The flows of the other particles only contain momentum-independent tadpole contributions, which are absorbed in the counter-term action, not shown in this figure. The dynamical part of the pion tadpole contains the $\pi\to 3\pi$ scattering, the lowest-lying scattering channel of the pions. Therefore, this tadpole contribution changes the result qualitatively. \hspace*{\fill}}
		\label{fig:2pt_flow}
	\end{minipage}
\end{figure}

Apart from the four-point vertex, we approximate mesonic vertices in \Cref{fig:2pt_flow} as (momentum-independent) derivatives of the effective potential $V$ on the physical minimum $\rho_0$. Note that the first and second derivatives of the potential are directly related to the mesonic propagators and given by
\begin{align}
	V'(\rho_0) &= Z_\phi^{-1}\Gamma^{(2)}_\pi(p=0)- k^2\nonumber\\
	V'(\rho_0) + 2\rho_0 V''(\rho_0) &=Z_\phi^{-1}\Gamma^{(2)}_\sigma(p=0)- k^2\,.
	\label{eq:PotentialFromSpecFunc}
\end{align} 
This translates the flowing renormalisation conditions \labelcref{eq:RenormCondMesons} into conditions for the effective potential. For more details, see~\Cref{sec:PotentialRegularisation}. We hasten to add that this by no means removes information from the potential, but merely provides a way to compute it consistently with the spectral functions. Its flow equation is derived from that of the effective action shown in~\Cref{fig:FlowofGamma} by evaluating it for constant fields $\phi_c$ and dividing out the space-time volume ${\cal V}$. 
\begin{align} 
	\partial_t V(\rho_c) =\frac{1}{{\cal V}} \partial_t \Gamma[q=0,\bar{q}=0,\phi_c]\,. 
\label{eq:FlowofV} 
\end{align} 

To solve \labelcref{eq:FlowofV}, we use the well-established fact that one can reformulate the flow of the potential as an advection-diffusion type partial differential equation and thus utilise methods from numerical hydrodynamics, see \cite{Grossi:2019urj, Grossi:2021ksl, Koenigstein:2021syz, Koenigstein:2021rxj, Steil:2021cbu, Ihssen:2022xkr, Ihssen:2023qaq, Ihssen:2023xlp, Zorbach:2024rre, Koenigstein:2025sse}.
In this work, we use the DiFfRG computational framework \cite{Sattler:2024ozv} to solve $V(\rho)$ with a finite-element method. 

In our computation, we go beyond the LPA approximation for the effective potential and incorporate the non-trivial momentum dependence of the two-point functions. For this purpose, we use an approach developed in \cite{Helmboldt:2014iya, IKPS2026}, where the two-point functions are split into a momentum-dependent and a field-dependent part. For the mesons, we get
\begin{align}\label{eq:PotentialPropTrunc}
	\Gamma^{(2)}_\pi[\rho](p)&=\Gamma^{(2)}_{\pi,\text{dyn}}(p) + Z_\phi(0) V'(\rho) + R_\phi\,,\\[1ex]\nonumber 
	\Gamma^{(2)}_\sigma[\rho](p)&=\Gamma^{(2)}_{\sigma,\text{dyn}}(p) + Z_\phi(0)\left(V'(\rho) + 2\rho V''(\rho)\right) + R_\phi\,,
\end{align}
where the dynamical part is given by the momentum dependence of the two-point function on the physical minimum $\rho_0$,
\begin{align}
	\Gamma^{(2)}_{\phi,\text{dyn}}(p)&=\Gamma^{(2)}_{\phi}[\rho_0](p)-\Gamma^{(2)}_{\phi}[\rho_0](0)\,.
	\label{eq:dynamical_two-point}
\end{align}
For the field-dependence of the quark propagator we use a different approximation. While the momentum dependence is again given by the one on the physical minimum, we assume the wave function to be field-independent and approximate the mass function by its mean-field behaviour, i.e., linearly in the field:
\begin{align}
	M(p,\rho) = M(p,\rho_0)\frac{\phi}{\phi_0}\,.
\end{align}

One difficulty that arises in this context is that the CS-cutoff makes the pole in the Wetterich equation responsible for convexity restoration (i.e. the convexity bound at $V''\to -k^2$) an integrable singularity, which leads to convexity restoration being numerically inaccessible. 

To remedy this problem, we use the \textit{CS-variant} regulator, see~\Cref{sec:PotentialRegulator}. This regulator combines the UV asymptotic of the CS-regulator with the convexity restoring IR behaviour of the polynomial-exponential regulator introduced in~\cite{Ihssen:2024miv}.

\begin{figure*}
	\centering
	\begin{minipage}{0.99\linewidth}
		\centering
		\begin{subfigure}[t]{0.48\linewidth}
			\includegraphics[width=\textwidth]{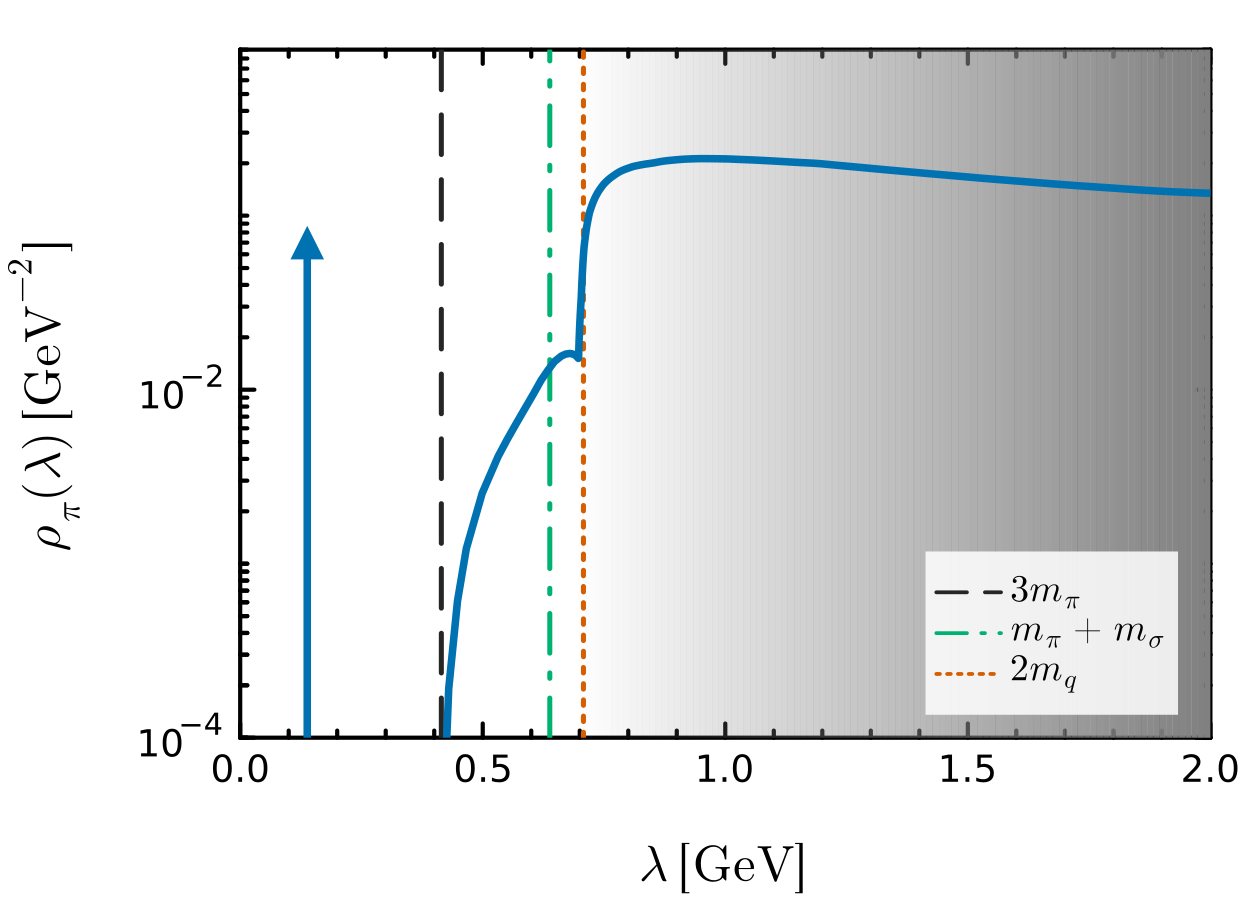}
			\caption{Pion spectral function. \hspace*{\fill}}
			\label{fig:rho_Pion}
		\end{subfigure}\hfill
		\begin{subfigure}[t]{0.48\linewidth}
			\includegraphics[width=\textwidth]{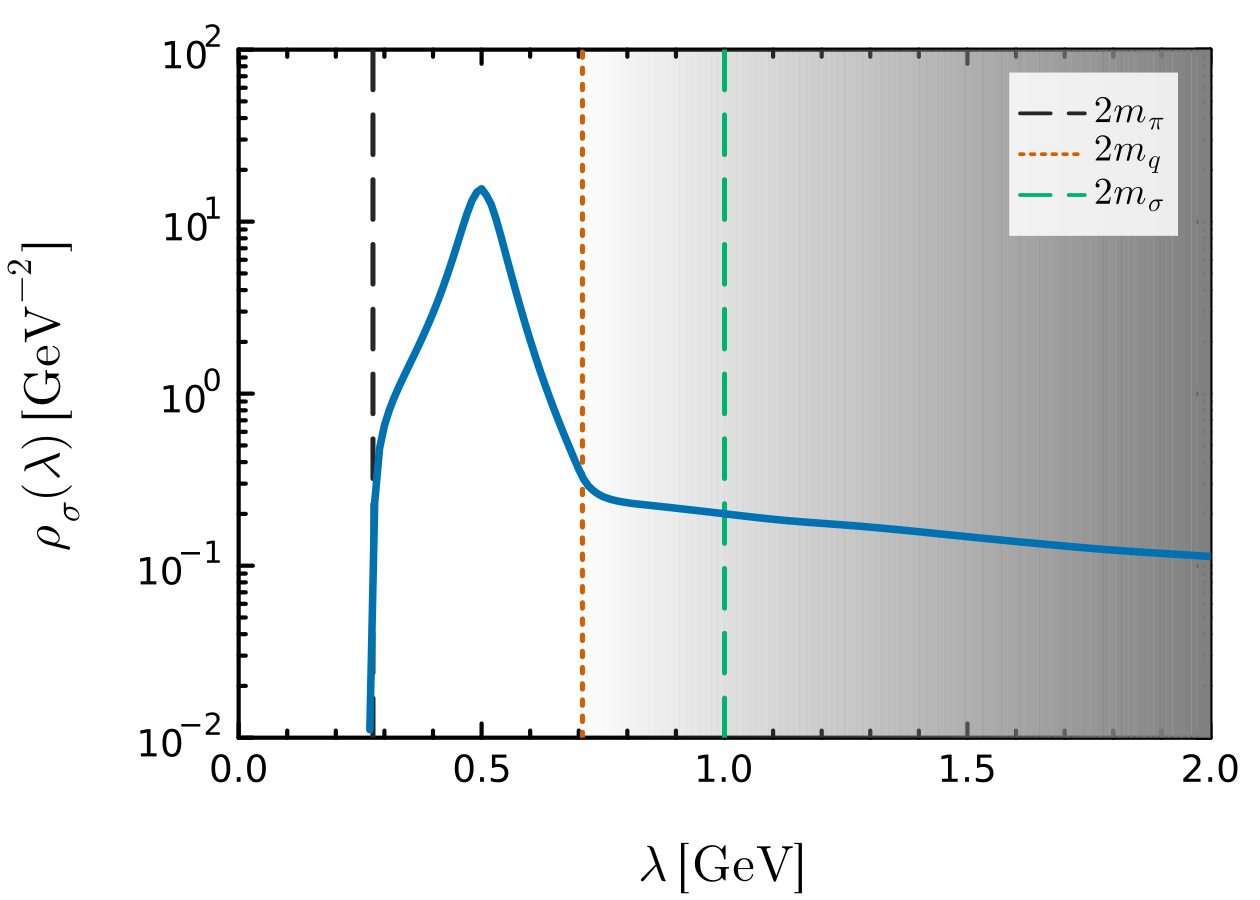}
			\caption{Sigma spectral function. \hspace*{\fill}}
			\label{fig:rho_Sigma}
		\end{subfigure}
		\caption{Spectral functions of the pion (left) and sigma (right). The leading scattering thresholds are indicated by dashed lines: for the pion, $\pi\to 3\pi$, $\pi\to\pi+\sigma$, and $\pi\to q+\bar{q}$; for the sigma, $\sigma\to2\pi$, $\sigma\to q\bar{q}$ and $\sigma\to 2\sigma$. Thresholds involving only stable asymptotic states produce sharp onsets, while those involving an intermediate sigma mode - namely $\pi\to\pi+\sigma$ and $\sigma\to2\sigma$ - are broadened due to the finite width of the sigma resonance. Thresholds of higher-order scattering processes are further suppressed and not distinguishable in the plots but present in the data. In the regime $\lambda > 1.5$\,GeV, residual RG-inconsistency of the one-loop initial condition leads to initial-scale artefacts in the spectral functions in the vicinity of the threshold $2 m_{q,\Lambda}$ with $\Lambda=0.7$\,GeV, for more details see \Cref{fig:rho_Mesons_RGcons} in \Cref{app:RG-consistency}. Effects from further, higher order thresholds at $k=\Lambda$ are negligible. 
		\hspace*{\fill} } 
		\label{fig:rho_Mesons_Vacuum}
	\end{minipage}
\end{figure*}
The combined system of spectral functions and the effective potential is solved iteratively, starting from a solution of the effective potential in the LPA approximation, which is then fed into the flow of the spectral functions. Once the solution for the spectral functions is obtained via fixed point iteration, the corresponding momentum dependencies are fed back into the flow of the effective potential as described above, and the process is repeated until full convergence is reached. 

For further details on the regularisation procedure and the numerical implementation, see~\Cref{subsec:Potential_flow}.

\section{Results}
\label{sec:Results}

We discuss results for the spectral functions of quarks and mesons ($\sigma$-mode and pions) in \Cref{sec:ResultsSpecFun}. These are the main computational results of the present work. The spectral functions are computed self-consistently and include all-order scattering effects of quarks, $\sigma$ and pions. In \Cref{sec:ResultsPotential} we discuss the results for the full effective potential that also includes all-order scattering effects.

\subsection{Spectral functions}
\label{sec:ResultsSpecFun}

The spectral functions of the scalar-pseudoscalar channel fields $\sigma$ and pions are discussed in 
\Cref{sec:ResultsMesons} and \Cref{sec:ResultsQuarks} respectively. The results for the spectral functions are shown in~\Cref{fig:rho_Mesons_Vacuum} and~\Cref{fig:rho_Quark_Vacuum}.

\subsubsection{Mesons}
\label{sec:ResultsMesons}

The two mesonic spectral functions are shown in \Cref{fig:rho_Mesons_Vacuum} and differ qualitatively in the nature of their one-particle contributions. The pion spectral function displays a sharp delta-function pole at $\lambda=m_\pi$, reflecting the stability of the pion as an asymptotic state. In contrast, the sigma spectral function exhibits a broad Breit-Wigner resonance peak rather than an isolated pole. This broadening is a direct consequence of the sigma's instability with respect to the decay $\sigma\to2\pi$: the two-pion threshold opens a cut at $2m_\pi$, which moves the pole off the real axis onto the second Riemann sheet.

The scattering continuum encodes the kinematic thresholds of accessible multi-particle channels, and the sharpness of each onset depends on whether the intermediate states are stable. Channels involving only stable asymptotic states - in the quark-meson model given by the quarks and pions - produce sharp, non-differentiable onsets in the spectral tail, while those involving the unstable sigma mode yield smoothed, broadened structures. The relevant thresholds are indicated by dashed lines in \Cref{fig:rho_Mesons_Vacuum}.

For the sigma, the two clearly identifiable and sharp onsets are $\sigma\to2\pi$ and $\sigma\to q+\bar{q}$. The contribution from any intermediate sigma propagator is correspondingly smeared and less prominent. 

The pion continuum has a richer structure. The dominant feature is the sharp onset of the $\pi\to q+\bar{q}$ channel at $2m_q$, which carries the bulk of the spectral weight in the high-energy tail. Below this threshold, the $\pi\to\pi+\sigma$ channel is the dominant contribution; however, since it involves an intermediate sigma, its onset is broadened and lacks a sharp threshold signature. The $\pi\to3\pi$ threshold at $3m_\pi$, while subdominant, is clearly identifiable as a smooth onset.

This $\pi\to 3\pi$ onset is of particular physical and methodological significance. First, it marks the true onset of the pion scattering continuum — the lowest-energy multi-particle state into which a pion can scatter. Second, and crucially, it corresponds to a physically realisable process: unlike quarks, pions are asymptotic states in QCD, so this channel has direct experimental relevance. Third, from a technical point of view, this contribution does not arise from the polarisation-type diagrams in the flow equation for $\Gamma^{(2)}_\pi$, but rather from the pion tadpole diagram, see \Cref{fig:2pt_flow}. An expansion to two-loop order reveals that the momentum dependence of the tadpole generates a sunset topology responsible for this threshold. This also explains its relative suppression compared to the $\pi\to q+\bar{q}$ and $\pi\to\pi+\sigma$ continuum, as these already arise at one-loop order. In the chiral limit it is precisely this $\pi\to3\pi$ channel that becomes gapless and dominates the infrared physics \cite{Kockler:2025kdt}.

Capturing this onset requires a momentum-dependent four-point vertex, as constructed in~\Cref{sec:AdditionalFourPointFunction}. In truncations with momentum-independent vertices, such as LPA or LPA', the tadpole contributes only a momentum independent term and therefore cannot generate a threshold structure of this kind. The correct description of the $\pi\to3\pi$ onset thus represents a systematic improvement over previous spectral function studies within the Quark-Meson and O(4) models~\cite{Tripolt:2013jra, Tripolt:2014wra, Tripolt:2016cey, Jung:2021ipc, Pawlowski:2017gxj, Topfel:2024iop}, where this channel was not accessible. 

As a numerical consistency check, we evaluated the mesonic spectral sum rule~\labelcref{eq:SpecSumRule}. For the full computation we find deviations of about $0.3\%$ in the pion channel and about $12\%$ in the sigma channel; the detailed comparison of different truncations is given in \Cref{app:KL-Existence}. The larger sigma deviation is traced back to the numerical treatment of higher-dimensional spectral integrals involving sharply peaked spectral functions, in particular in the $\sigma\to2\sigma$ contribution. We therefore use the sum-rule violation as a lower estimate of the numerical error.

\subsubsection{Quarks}
\label{sec:ResultsQuarks}

\begin{figure*}
	\centering
	\begin{minipage}{0.99\linewidth}
		\centering
		\begin{subfigure}[t]{0.48\linewidth}
			\includegraphics[width=\textwidth]{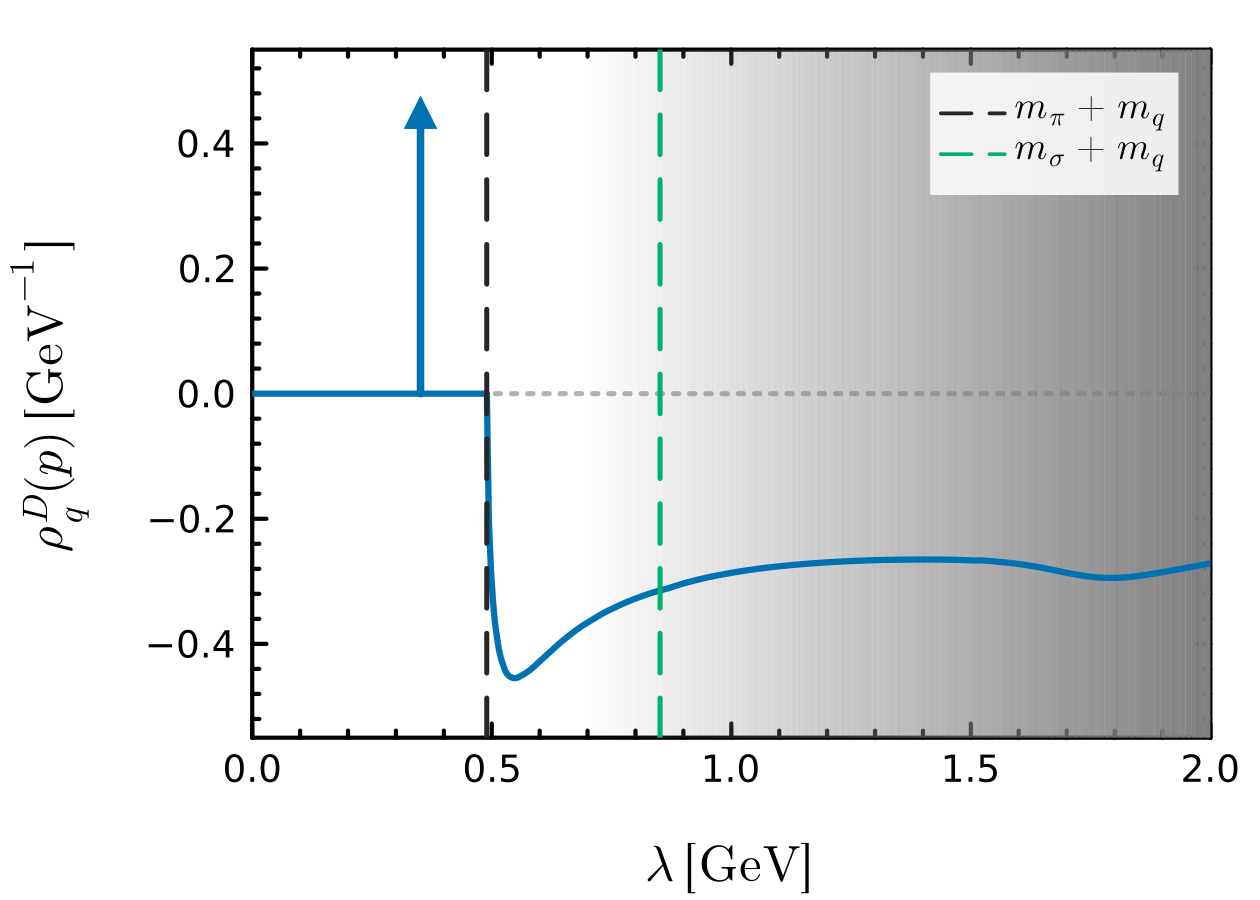}
			\caption{Quark Dirac spectral function. \hspace*{\fill}}
			\label{fig:rho_Quark_dirac}
		\end{subfigure}\hfill
		\begin{subfigure}[t]{0.48\linewidth}
			\includegraphics[width=\textwidth]{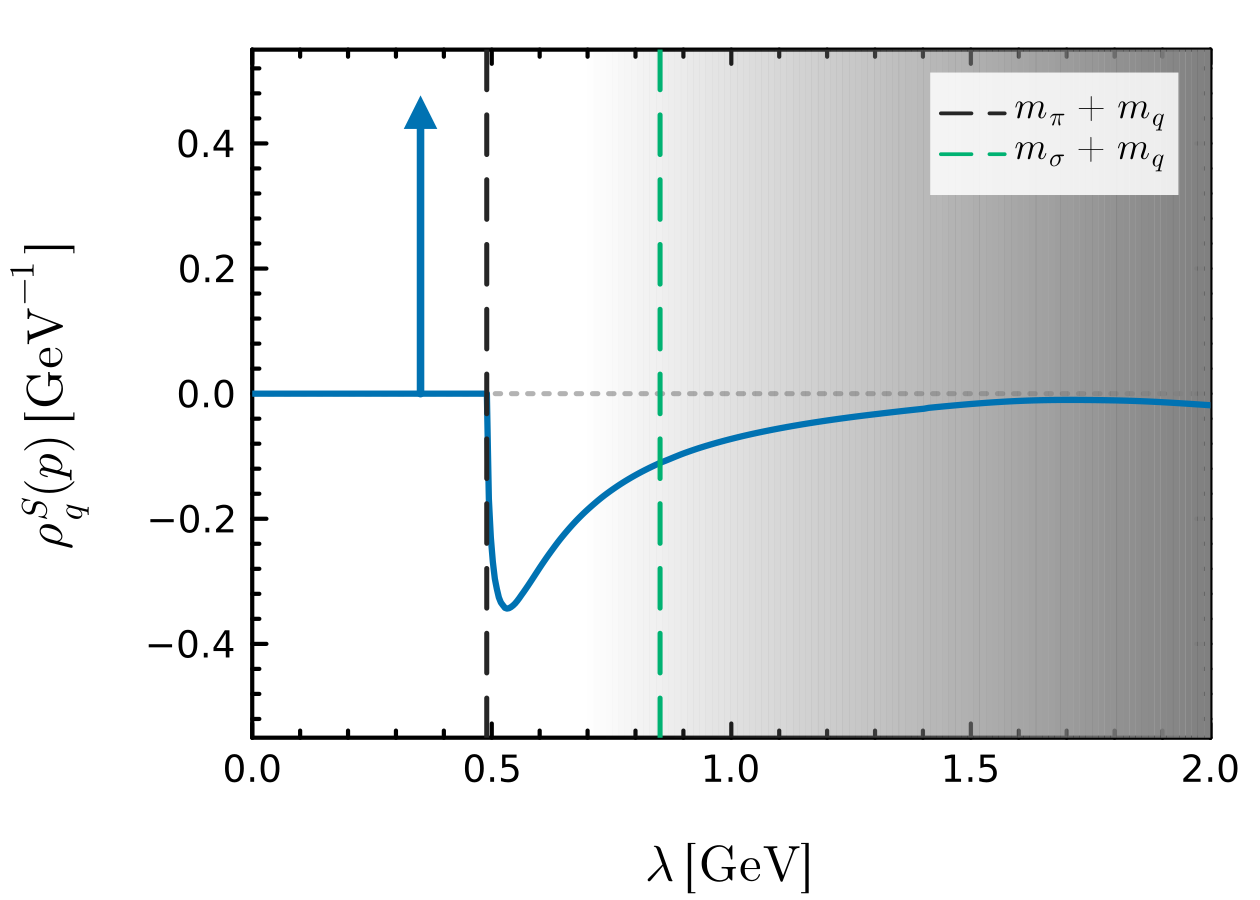}
			\caption{Quark scalar spectral function. \hspace*{\fill}}
			\label{fig:rho_quark_scalar}
		\end{subfigure}
		\caption{
			Spectral functions of the Dirac component $\rho_q^D$ (left) and scalar component $\rho_q^S$ (right) of the quark propagator. Both components share the same pole position and threshold structure. The two leading scattering thresholds, $q\to q+\pi$ and $q\to q+\sigma$, are indicated by dashed lines. The pion threshold produces a sharp onset and dominates the spectral weight. The sigma threshold is broadened by the finite width of the sigma resonance and is subleading, making it difficult to identify as a distinct feature. In the regime $\lambda > 1.5$\,GeV, residual RG-inconsistency of the one-loop initial condition leads to initial-scale artefacts in the spectral functions in the vicinity of the thresholds $m_{q,\Lambda}+m_{\pi,\Lambda}$ with $\Lambda=0.7$\,GeV, for more details see \Cref{fig:rho_Quark_RGcons} in \Cref{app:RG-consistency}. Effects from further, higher order thresholds at $k=\Lambda$ are negligible. 			
			\hspace*{\fill} } 
		\label{fig:rho_Quark_Vacuum}
	\end{minipage}
\end{figure*}

The full quark propagator admits a decomposition into two independent Dirac tensor structures, as detailed in \Cref{eq:QuarkPropDecomposition}, each associated with a spectral function defined via \Cref{eq:rhoQuarkFromG}. Although the two components $G_D$ and $G_S$ carry different canonical mass dimensions, the corresponding spectral functions can be brought to a common dimension by multiplying the Dirac component by the spectral parameter $\lambda$, as indicated in \Cref{eq:rhoQuarkFromG}. The resulting spectral representations are displayed in \Cref{fig:rho_Quark_Vacuum}.

The two components share identical analytic structure: both exhibit the same one-particle pole position and the same branch cut support. This is a direct consequence of the fact that $G_D$ and $G_S$ are related components of a single propagator, so their singularities are necessarily tied to the same physical states. The spectral functions therefore differ only in their overall quantitative structures, not in the location of any threshold or pole.

Within the Quark-Meson model, the quark scattering continuum receives contributions from two leading channels: $q\to q+\pi$ and $q\to q+\sigma$. The pion channel produces a sharp, clearly visible onset in the spectral tail, consistent with the pion being a stable asymptotic state. The sigma channel, by contrast, yields a broadened onset for the same reason as discussed in the meson sector: the finite width of the sigma smears the corresponding threshold. The smearing, in addition to the onset being subleading in this spectral range, leads to the $q\to q+\sigma$ onset not being clearly distinguished. The positions of both onsets are indicated in \Cref{fig:rho_Quark_Vacuum}.

It is instructive to contrast these results with quark spectral functions computed in full QCD, where gluonic degrees of freedom are present. Such computations have been carried out using Dyson-Schwinger equations in~\cite{Pawlowski:2024kxc}. In full QCD, the quark still exhibits a one-particle peak, see also \cite{Wieland:2026iml,Alkofer:2026zhf}, but the scattering continuum sets in immediately at the pole, leaving no gap between the pole and the onset of the spectral tail. This qualitative difference has a transparent origin: in QCD, the leading scattering channel is $q\to q+A$, where $A$ denotes the gluon, along with higher-order multi-parton processes. Since the gluon spectral function has support at arbitrarily small spectral values - reflecting its confinement-related infrared structure - the corresponding threshold in the quark spectral function is pushed all the way down to the pole itself. In the Quark-Meson model, by contrast, the lightest exchanged boson is the pion with a finite mass, which sets a gap between the quark pole and the onset of inelastic scattering.

As for the mesons, we checked the quark spectral sum rule associated with the large-momentum normalisation of the Dirac component in \labelcref{eq:Quark-KL}. For the full computation we find a deviation of about $6\%$. This is smaller than the sigma-channel deviation but larger than the pion one, and we include it in the numerical error estimate discussed in \Cref{app:KL-Existence}.

\subsection{Effective potential}
\label{sec:ResultsPotential}

We now turn to the effective potential computed with momentum-dependent propagators as input. The respective setup is outlined in \Cref{subsec:Potential_flow}. The flow equation for the effective potential receives two types of input derived from the spectral functions: first, the $k$- and $p$-dependent Euclidean propagators of quarks and mesons, which encode the full momentum dependence of the two-point functions on the physical minimum; 
second, the scale-dependent trajectories of the first two potential derivatives evaluated on the equation of motion, as defined in~\Cref{eq:PotentialFromSpecFunc}, which serve as renormalisation conditions for the field-dependent part of the potential. Together, these inputs allow for a consistent incorporation of the full propagator momentum dependence on the physical minimum $\rho_0$ into the potential flow, going beyond the local potential approximation.

\begin{figure*}
	\centering
	\begin{minipage}{0.99\linewidth}
		\centering
		\begin{subfigure}[t]{0.45\linewidth}
			\includegraphics[width=\textwidth]{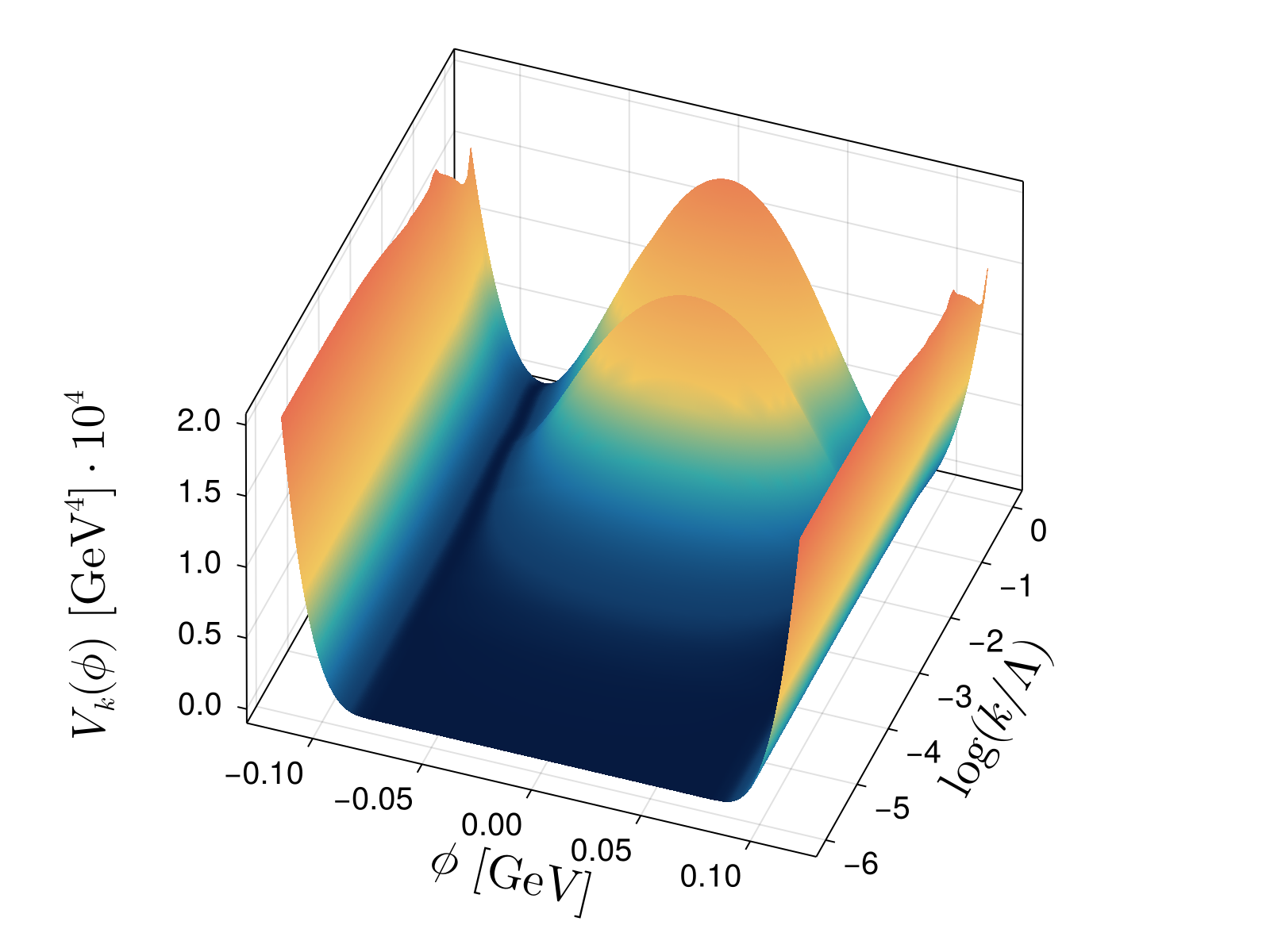}
			\caption{Scale- and field-dependent effective potential. \hspace*{\fill}}
		\end{subfigure}\hfill
		\begin{subfigure}[t]{0.45\linewidth}
			\includegraphics[width=\textwidth]{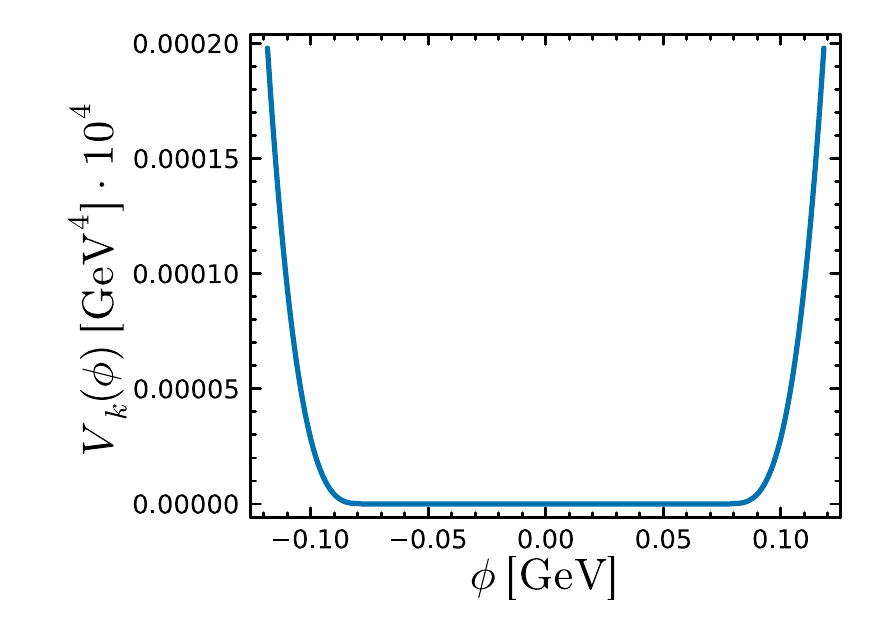}
			\caption{Effective potential at $k=0$. \hspace*{\fill}}
		\end{subfigure}
		\caption{Effective potential, both during the flow and at $k=0$. The absolute value of the potential is fixed to be zero in the chiral minimum. \hspace*{\fill} } 
		\label{fig:PotentialResult}
	\end{minipage}
\end{figure*}

We show the resulting effective potential in~\Cref{fig:PotentialResult}, indicating the position of $\rho_0$. Clearly visible is the gradual development of the flat, non-physical region of the effective potential, resulting in a non-analytic, convex potential at the final scale.

\begin{figure*}
\centering
\begin{minipage}{0.99\linewidth}
	\centering
	\begin{subfigure}[t]{0.32\linewidth}
		\includegraphics[width=\textwidth]{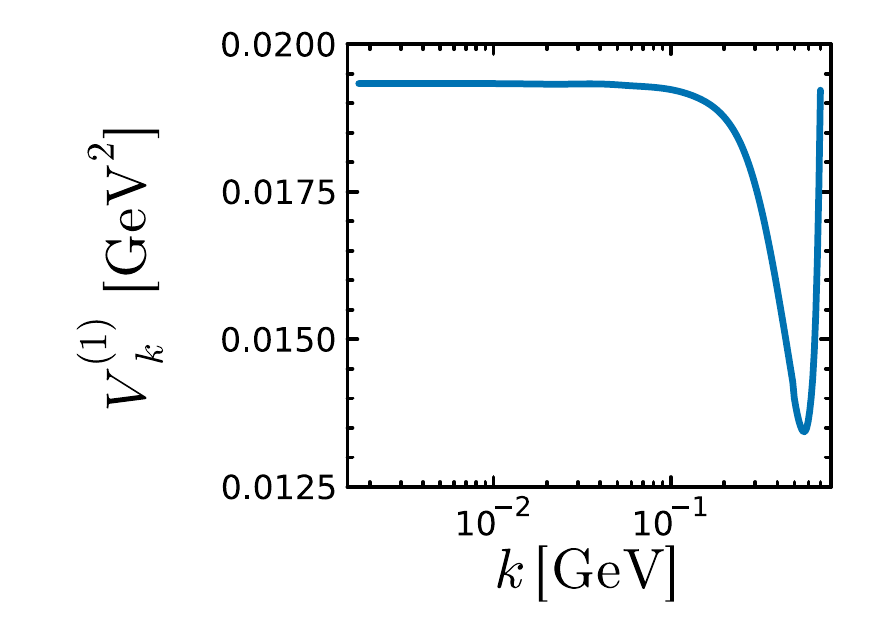}
		\label{fig:d1V}
	\end{subfigure}
	\begin{subfigure}[t]{0.32\linewidth}
		\includegraphics[width=\textwidth]{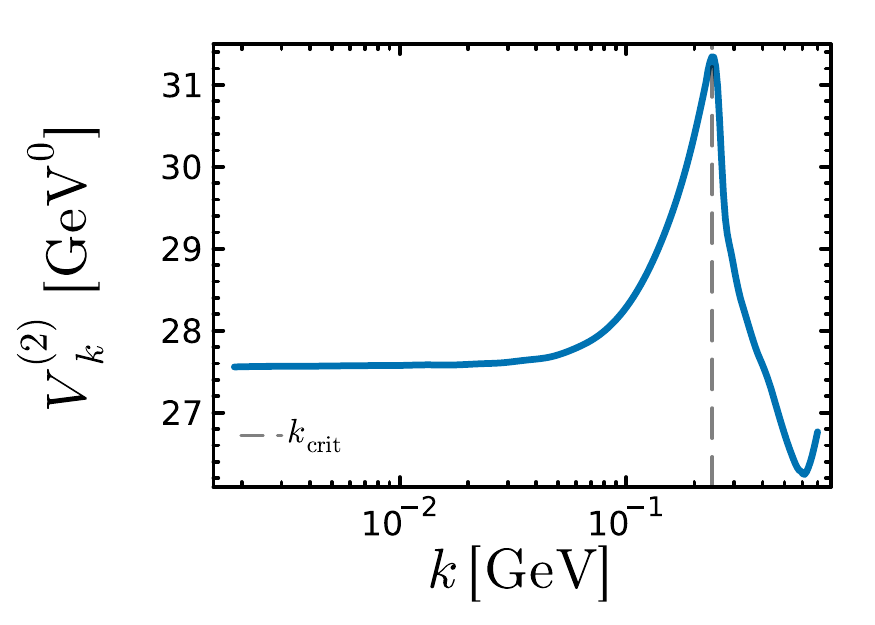}
		\label{fig:d2V}
	\end{subfigure}
	\begin{subfigure}[t]{0.32\linewidth}
		\includegraphics[width=\textwidth]{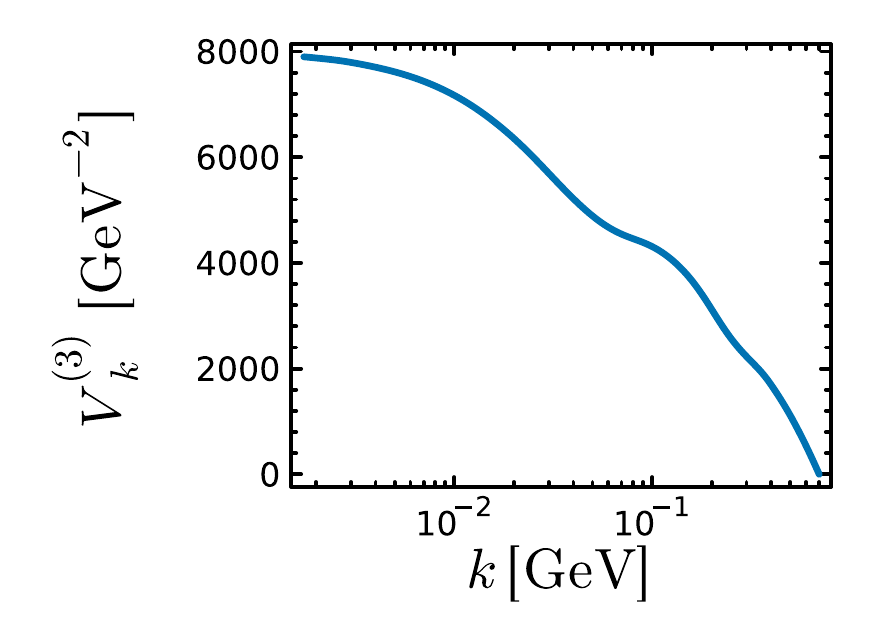}
		\label{fig:d3V}
	\end{subfigure}
	\hfill
	\caption{Scale evolution of the first three derivatives of the effective potential evaluated on the physical minimum $\rho_0$. The first and second derivatives are fixed by the renormalisation conditions; the third constitutes genuine output of the potential flow. In the second and third derivative, the most visible feature is the point $k_{\textrm{crit}}\approx 0.24\,\mathrm{GeV}$, at which the $\sigma$ becomes unstable, i.e. its pole mass becomes smaller than twice the pion pole mass.
		\hspace*{\fill} } 
	\label{fig:PotentialDerivatives}
\end{minipage}
\end{figure*}

Since the physical theory is defined on the equation of motion, only the behaviour of the potential and its derivatives at $\rho_0$ enter directly into observables. Higher-order interaction vertices are determined by the corresponding higher derivatives of the potential on the minimum, and processes involving the sigma mode are particularly sensitive to these. The first three derivatives of the effective potential are displayed in~\Cref{fig:PotentialDerivatives}. The first and second derivatives are uniquely fixed by the renormalisation conditions and the momentum-dependent flows of the inverse propagators. The third derivative, by contrast, is a genuine output of the potential flow and thus carries non-trivial information about higher-order mesonic interactions beyond what is encoded in the two-point functions.

The most pronounced feature of the second and third derivative is the peak at $k_{\mathrm{crit}}\approx 0.24\,\mathrm{GeV}$. At this point, the scattering onset of the $\sigma$, located at twice the pion pole mass, crosses the $\sigma$ pole. Consequently, this pushes the latter onto the second Riemann sheet. The $\sigma$ is no longer stable and its one-particle delta pole broadens into a Breit-Wigner shape.

%

\section{Conclusion}
\label{sec:Conclusions}

We have computed spectral functions of quarks and mesonic composites in low-energy QCD. The computation was done in a two-flavour Quark-Meson model which emerges from full QCD via the decoupling of gluonic degrees of freedom. The present functional renormalisation group analysis goes qualitatively beyond previous spectral fRG computations in the QM-model and (non-)linear sigma models, e.g.~\cite{Tripolt:2021jtp,Heller:2021wan, Tan:2021zid, Jung:2021ipc, Roth:2021nrd, Horak:2022aza, Horak:2022myj, Roth:2023wbp, Horak:2023hkp, Roth:2024rbi, Chen:2024lzz, Fu:2024rto, Pawlowski:2024kxc, Topfel:2024iop, Roth:2025hcm, Kockler:2025kdt, Tan:2025bsv, Roth:2026zrs} in three qualitative aspects:\\[-1ex] 

First, the spectral functions are fed back into the flow self-consistently, generating all-order decays. Together with the inclusion of a momentum-dependent four-pion vertex obtained from its Bethe-Salpeter equation this gives access to all-order scattering events. Specifically, we have resolved $\pi\to 3\pi$ scatterings, the lowest kinematically accessible channel for the pion which has been absent so far. \\[-2ex] 

Second, rescattering events of quarks and mesonic composites have been taken into account with the fRG approach with emergent composites. Only with these events the full scattering dynamics of quarks and mesons is accessible. \\[-2ex]

Third, mesonic self-scattering events are included to all orders through a full effective potential. In other spectral computations these scattering events have only been considered to a relatively low order. \\[-1ex]

In combination, these novel developments give quantitative access to the spectral properties of low-energy QCD, and specifically allow us to resolve regimes with soft modes where multiscattering and decays become important. 

We proceed with a short review of the computational results, more details can be found in \Cref{sec:Results}. The pion spectral function exhibits a stable one-particle pole and the $\pi\to 3\pi$ threshold marks the onset of the scattering continuum, see \Cref{fig:rho_Pion}. The spectral function of the sigma-mode resembles a Breit-Wigner resonance, reflecting its instability towards a decay into two pions, see \Cref{fig:rho_Sigma}. For the quarks, both Dirac tensor structures of the propagator are resolved, displaying a clear separation between the one-particle pole and the scattering onset at $m_q+m_\pi$, see \Cref{fig:rho_Quark_Vacuum}. We note in passing that in Landau gauge QCD the spectral gluon features a cut starting at $p=0$, and we expect the full quark spectral function to carry spectral weight below the $q\to q +\pi$ scattering threshold, see \cite{Horak:2022aza, Pawlowski:2024kxc}. 

Mesonic higher-order self-scatterings and quark-meson rescatterings become particularly important in the vicinity of the chiral crossover, where soft pion and sigma modes proliferate and multi-particle processes contribute significantly to spectral observables~\cite{Braun:2023qak}. The present work therefore lays the foundation for future realtime studies of soft modes near the chiral phase transition in low-energy QCD in the vacuum and at finite temperature and density, in particular in the vicinity of the Critical End Point or rather the Onset of New Phases, \cite{Pawlowski:2025jpg}, see also the reviews \cite{Fischer:2026uni, Fischer:2026vkc}. One of the many direct applications of the present framework is the computation of transport coefficients in the phase structure with functional QCD. This application also uses gluonic spectral properties as analysed in functional QCD in \cite{Haas:2013hpa, Ilgenfritz:2017kkp, Cyrol:2018xeq, Horak:2021pfr, Horak:2021syv, Horak:2022myj, Horak:2023xfb}.

Specifically we aim at shear and bulk viscosities and heavy quark diffusion coefficient at finite temperature and density. We hope to report on progress in this direction in the near future.

%

\subsection*{Acknowledgements}

We thank Gernot Eichmann, Markus Huber, Jan Horak and Joannis Papavassiliou for discussions and collaboration on related projects. This work is done within the fQCD-collaboration~\cite{fQCD}, and we thank the members for discussion and collaborations on related projects. This work is funded by the Deutsche Forschungsgemeinschaft (DFG, German Research Foundation) under Germany's Excellence Strategy EXC 2181/1 - 390900948 (the Heidelberg STRUCTURES Excellence Cluster) and under the Collaborative Research Centre SFB 1225 (ISOQUANT). KK acknowledges funding by the GSI Helmholtzzentrum für Schwerionenforschung and by HGS-HIRe for FAIR. JW acknowledges funding by the DFG under grant numbers HU 2176/3-1.
FRS acknowledges funding by the GSI Helmholtzzentrum f\"ur Schwerionenforschung and by HGS-HIRe for FAIR.
FRS is supported by the Deutsche Forschungsgemeinschaft (DFG, German Research Foundation) through the Emmy Noether Programme Project No. 54526179.

\appendix

\section{Shift symmetry, flows and Dyson-Schwinger equations}
\label{app:Shift-DSE+fRG}

The flow equation \labelcref{eq:Ren-flow} has been obtained without using this shift and at the initial scale it is sourced by the quark loop $\partial_t \Gamma_q$: For momentum-independent regulators it is a function of $\phi'$, defined in \labelcref{eq:Shiftphi}, and $\partial_t R_q$ with $R_q=0$,
\begin{align} \nonumber 
	\partial_t \Gamma_q=&\,- \Tr \partial_t R_q \,G_q[\Phi;R_q,R_\phi] \\[1ex]
	=& \,- \Tr \partial_t R_q \,G_q[\Phi';0,R_\phi] \,.
	\label{eq:q-FlowFlow'}
\end{align}
where $\Phi'=(q,\bar q,\phi')$ and $\phi'$ is the shifted field \labelcref{eq:Shiftphi}. Moreover, for the sake of simplicity we dropped the flowing counter term action $\partial_t S_\textrm{ct}$. 

The initial effective action is quadratic in the field $\phi$, making the initial meson propagator (and consequently the mesonic contribution to the flow) field independent. Thus, the field dependence gets sourced from the quark loop. As this is a function of $\phi'$, not $\phi$, it can only source this kind of dependence. Accordingly, the full effective action computed from the initial effective action and the flow \labelcref{eq:Ren-flow} satisfies the relation 
\begin{align} \nonumber 
	\Gamma_{k}[\Phi;R_q,R_\phi]= &\, \Gamma_{k}[\Phi';0,R_\phi]\\[1ex] 
	& -\frac12 \int_x 	
	\textrm{tr}\,\left(c_\phi\, \phi'\right)+{\cal C}_k \,, 
	\label{eq:ShiftSymFromFlow}
\end{align} 
with $k$-dependent constant ${\cal C}_k$ and $\phi'$ defined in \labelcref{eq:Shiftphi}, 
\begin{align} 
	\partial_t \phi' =\frac{1}{h_\phi} \partial_t R_q \,.
	\label{eq:dotphi'}
\end{align}
Note that this argument extends to the RG-consistent term,
\begin{align} 
	\Delta\Gamma^\textrm{\tiny{RGC}}_{\Lambda}=\Delta\Gamma^\textrm{\tiny{RGC}}_{\Lambda}[\Phi';S_{\text{QM}}]\,,
\end{align}
as it is generated by the flow. Now we use the mesonic Dyson-Schwinger equation for the QM-model with the classical action \labelcref{eq:Scl_QM}, adding the regulator terms, 
\begin{align} \nonumber 
	\frac{\delta \Gamma_{k}} {\delta\phi}[\Phi;R_q,R_\phi] = &\,\left\langle 	\frac{\delta S_{\textrm{QM}}} {\delta \phi}[\hat \Phi] \right\rangle \\[2ex]\nonumber 
	&\hspace{-2.5cm} =- h_\phi \tau \,G_q[\Phi;R_q,R_\phi] +	\frac{\delta S_{\textrm{QM}}} {\delta \phi}[ \Phi] \,, 
	\\[2ex]
	&\hspace{-2.5cm} =- h_\phi \tau \,G_q[\Phi';0,R_\phi] +	\frac{\delta S_{\textrm{QM}}} {\delta \phi}[ \Phi'] -\frac12 c_\phi\,, 
	\label{eq:DSE-QMModel}
\end{align} 
where the hat indicates the field operators. \Cref{eq:DSE-QMModel} entails that \labelcref{eq:q-FlowFlow'} is nothing but the mesonic DSE, projected on $\partial_t \phi'$ in \labelcref{eq:dotphi'}. We find 
\begin{align}
	\partial_t \Gamma_q= \int_x \partial_t \phi' \left(\frac{\delta \Gamma_{k}}{\delta \phi} -\frac{\delta S_{\textrm{QM}} } {\delta \phi}[ \Phi]\right)\,.
	\label{eq:q-FlowDSE}
\end{align}
With \labelcref{eq:q-FlowDSE}, the flow equation \labelcref{eq:Ren-flow} takes the form 
\begin{align} \nonumber 
	&\partial_t \Gamma_{k}[\Phi;R_q,R_\phi] - \int_x\partial_t \phi' \left(\frac{\delta \Gamma_{k}}{\delta \phi} -\frac{\delta S_{\textrm{QM}} } {\delta \phi}[ \Phi] \right)\\[1ex]
	=\,	& \frac12 \textrm{Tr} \,G_\phi[\Phi',R_\phi]\,\partial_t R_\phi\,.
	\label{eq:Flow+DSE}
\end{align}
Now we use \labelcref{eq:ShiftSymFromFlow} for the right-hand side of \labelcref{eq:Flow+DSE}, 
\begin{align} \nonumber 
	&\partial_t \Gamma_{k}[\Phi;R_q,R_\phi] - \int_x\partial_t \phi' \left(\frac{\delta \Gamma_{k}}{\delta \phi} -\frac{\delta S_{\textrm{QM}} } {\delta \phi}[ \Phi] \right)\\[1ex]\nonumber 
	= \,&\partial_t \Gamma_{k}[\Phi';0,R_\phi]- \int_x\partial_t \phi' \frac{\delta \Gamma_{k}[ \Phi']}{\delta \phi'} \\[1ex]
	& +\int_x \partial_t \phi' \frac{\delta S_{\textrm{QM}}[ \Phi']} {\delta \phi'}-\frac12 \int_x 	
	\textrm{tr}\,\left(c_\phi\, \partial_t\phi'\right)\,. 
	\label{eq:q-Flow-Flow'}
\end{align}
The second line is simply the $t$-derivative of the effective action at fixed $\Phi'$ and the third line is a field-independent (but $k$-dependent) constant. Now we relabel the free variable $\Phi'\to \Phi$ and arrive at 
\begin{align}
	\partial_t\Gamma_k = \frac{1}{2}\mathrm{Tr}\,G_\phi[\Phi;0,R_\phi]\dot{R}_\phi-\partial_t S_{\text{ct}}+\frac12 \int_x
	\textrm{tr}\,\left(\partial_t c_\phi\, \phi\right) \,, 
	\label{eq:RenShift-flowApp} 
\end{align}
which is nothing but \labelcref{eq:RenShift-flow}. In summary, we have shown the equivalence of \labelcref{eq:Ren-flow} and \labelcref{eq:RenShift-flowApp} (or \labelcref{eq:RenShift-flow}) by using the mesonic DSE that carries the shift symmetry of the path integral measure on the level of the effective action. 

\section{RG-consistency and initial condition}
\label{app:RG-consistency}

\begin{figure*}
	\centering
	\begin{minipage}{0.99\linewidth}
		\centering
		\begin{subfigure}[t]{0.48\linewidth}
			\includegraphics[width=\textwidth]{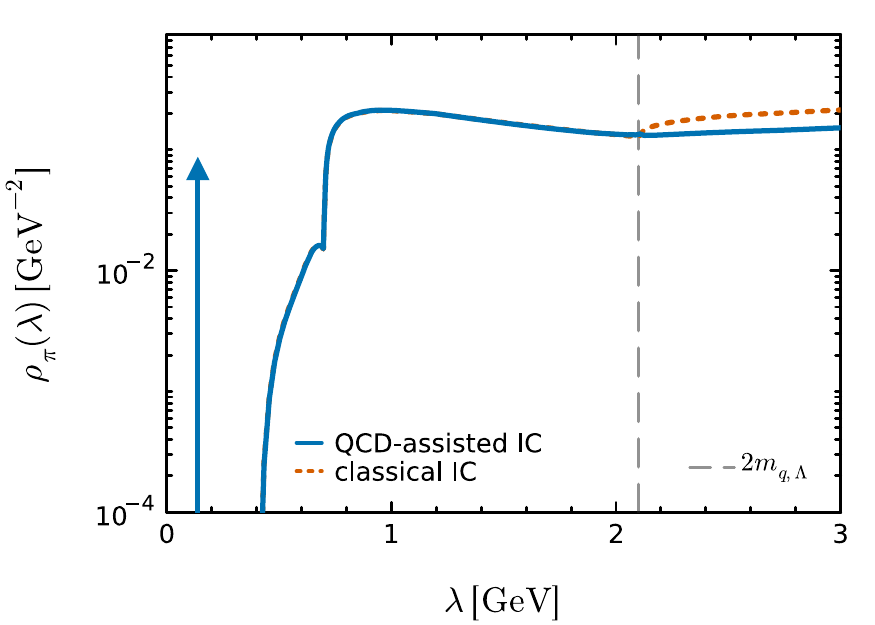}
			\caption{Pion spectral function. \hspace*{\fill}}
			\label{fig:rho_Pion_RGcons}
		\end{subfigure}\hfill
		\begin{subfigure}[t]{0.48\linewidth}
			\includegraphics[width=\textwidth]{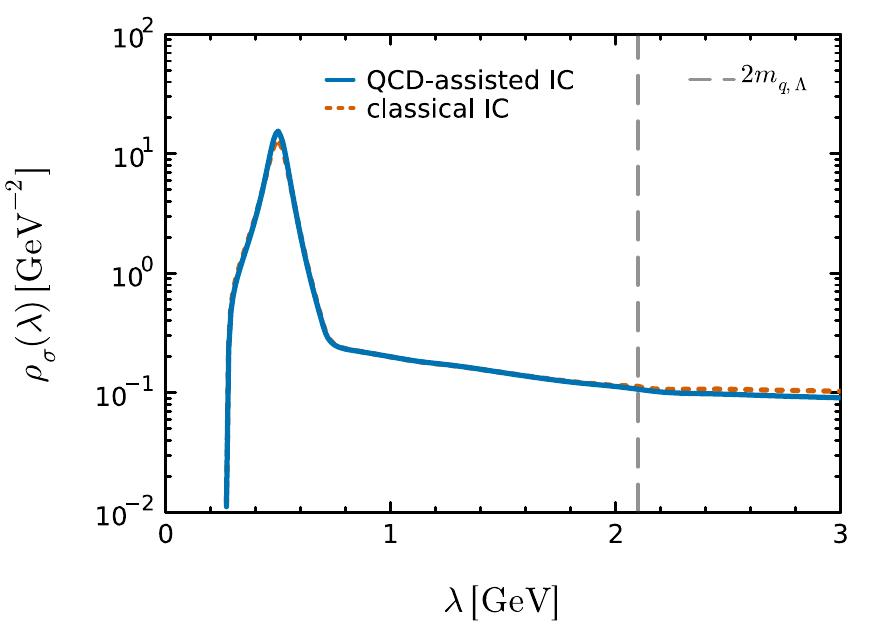}
			\caption{Sigma spectral function. \hspace*{\fill}}
			\label{fig:rho_Sigma_RGcons}
		\end{subfigure}
		\caption{Comparison of the mesonic spectral functions of an approximately RG-consistent QCD-assisted initial condition and with an RG-inconsistent classical initial condition. Above the initial threshold $2m_{q,\Lambda}$, they differ significantly, while below the difference is only marginal, due to the locality of the flow. Initial-scale artefacts at the initial threshold are present in both cases, but significantly reduced in the one-loop case.
		\hspace*{\fill} } 
		\label{fig:rho_Mesons_RGcons}
	\end{minipage}
\end{figure*}
\begin{figure*}
	\centering
	\begin{minipage}{0.99\linewidth}
		\centering
		\begin{subfigure}[t]{0.48\linewidth}
			\includegraphics[width=\textwidth]{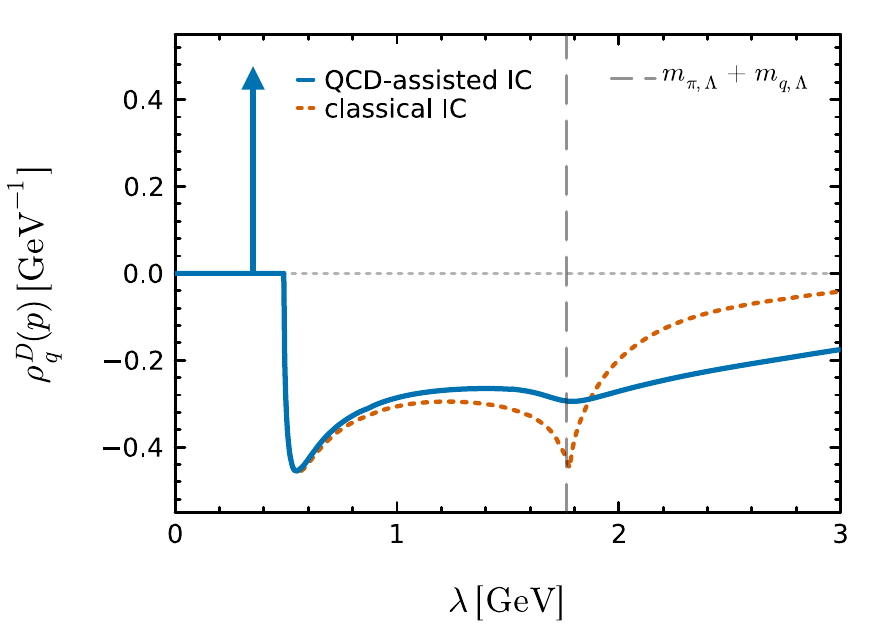}
			\caption{Quark Dirac spectral function. \hspace*{\fill}}
			\label{fig:rho_Quark_dirac_RGcons}
		\end{subfigure}\hfill
		\begin{subfigure}[t]{0.48\linewidth}
			\includegraphics[width=\textwidth]{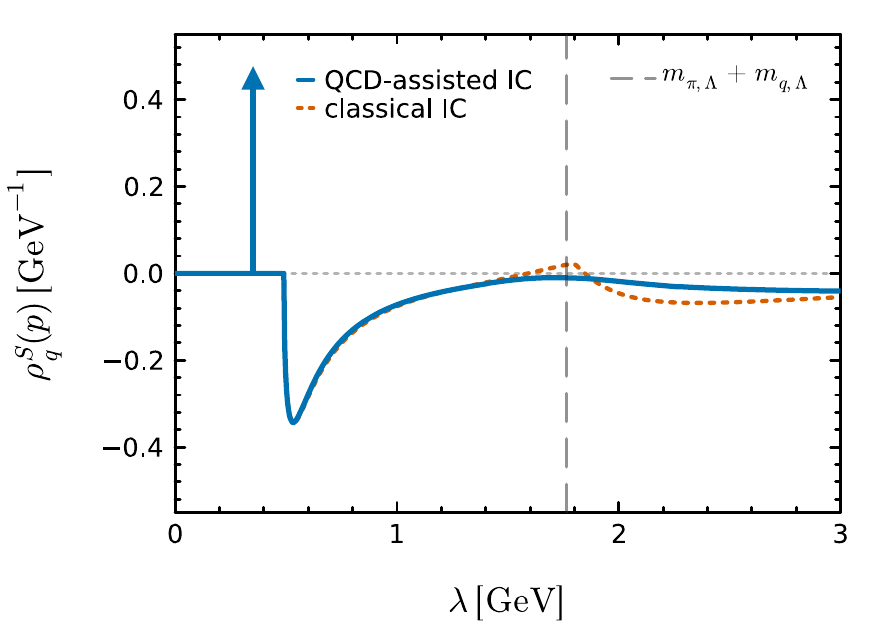}
			\caption{Quark scalar spectral function. \hspace*{\fill}}
			\label{fig:rho_quark_scalar_RGcons}
		\end{subfigure}
		\caption{
			Comparison of the quark spectral functions of an approximately RG-consistent QCD-assisted initial condition and with an RG-inconsistent classical initial condition. Compared to the mesonic spectral functions, the initial-scale artefacts are more pronounced and are significant even below the initial onset.
			\hspace*{\fill} } 
		\label{fig:rho_Quark_RGcons}
	\end{minipage}
\end{figure*}

In this section we go into more detail on the choice of initial condition and what effects it has on the resulting spectral function. More precisely, we compare an initial condition with classical momentum dependence to one with a perturbative one-loop behaviour.

The classical initial condition is obtained by setting $\Delta\Gamma^{(2)}_{\textrm{\tiny RGC},\Lambda}=\Delta\Gamma^{(2)}_{\textrm{\tiny RGC},\Lambda}(p=0)$, while the QCD-assisted initial condition uses the perturbative tail in~\labelcref{eq:RGC-1loop}. Thus, both computations have the same infrared renormalisation conditions, but differ in the momentum dependence already present at $k=\Lambda$.

The flow of the imaginary part of the inverse propagator has the property that its support is bounded from below by a (scale-dependent) onset $m_{\text{scat}, k}$,
\begin{align}
	\text{Im}\,\partial_t\Gamma^{(2)}(\omega<m_{\text{scat}, k}) = 0\,.
\end{align}
This onset is the onset of the scattering continuum at the scale $k$, i.e. the sum of the pole masses or thresholds of the lowest-lying scattering state. As this point marks the start of the branch cut on the real time axis, the flow is non-analytic at this point. Consequently, this non-analyticity translates to the two-point function. Depending on the dimension and the nature of the onset (the statistics and number of particles in the corresponding scattering state), this non-analyticity is a discontinuity, a sharp onset or a discontinuity of a higher derivative.

The real part of the two-point function can be obtained via the Kramers-Kronig relations. A discontinuity in the imaginary part for example corresponds to a logarithmic divergence in the real part, while a sharp onset corresponds to a kink.

As discussed in the main text, we do not feed the QCD-assisted initial condition into the flow, but add it as global correction. This introduces a slight inconsistency in the resulting two-point function. For fully RG-consistent initial conditions, the resulting two-point function would be smooth in momentum, apart from the physical scattering onsets. However, as our initial condition is not RG-consistent but resembles the a physical UV tail for the Quark-Meson model, the resulting two-point function still have non-analyticities at the initial onset $m_{\text{scat}, \Lambda}$.

In~\Cref{fig:rho_Mesons_RGcons,fig:rho_Quark_RGcons} we compare two choices of initial conditions at the example of the spectral functions of Quarks and mesons: an RG-inconsistent classical and an approximately RG-consistent QCD-assisted initial condition. 

The former has a sharp non-analyticity at the initial onset. Consequently, it shows pronounced initial-scale artefacts, especially at the initial onsets, which are indicated in grey. While QCD-assisted initial condition certainly improves the situation, kinks are still clearly visible and RG-consistency is violated.

This comparison further shows that the low-energy spectral structures discussed in the main text are not affected by the initial-scale artefact. The latter mainly affects the UV tail and the vicinity of the initial scattering onset, while the physical thresholds below this region are qualitatively stable under the replacement of the classical by the QCD-assisted initial condition. The mesonic spectral functions show only minimal differences, whereas the quark spectral functions are more sensitive to initial-scale artefacts, but still differ only quantitatively in the low-energy region.

Before we close this section, we want to mention one consequence of a one-loop initial action, where the decay of the Yukawa coupling in QCD is not taken into account. As the one-loop diagrams go with $p^2\log(p^2)$ and $\log(p^2)$ in the limit $p\to\infty$, the wave functions asymptotically go as
\begin{align}
	Z(p^2\to\infty)\sim \log(p^2)\;.
\end{align}
As this limit corresponds to the inverse spectral sum rule, i.e. the integral over the spectral functions, the latter cannot be fulfilled, and the spectral functions cannot be normalised to unity.

\begin{figure*}
	\centering
	\begin{minipage}{0.99\linewidth}
		\centering
		\begin{subfigure}[t]{0.45\linewidth}
			\includegraphics[width=\textwidth]{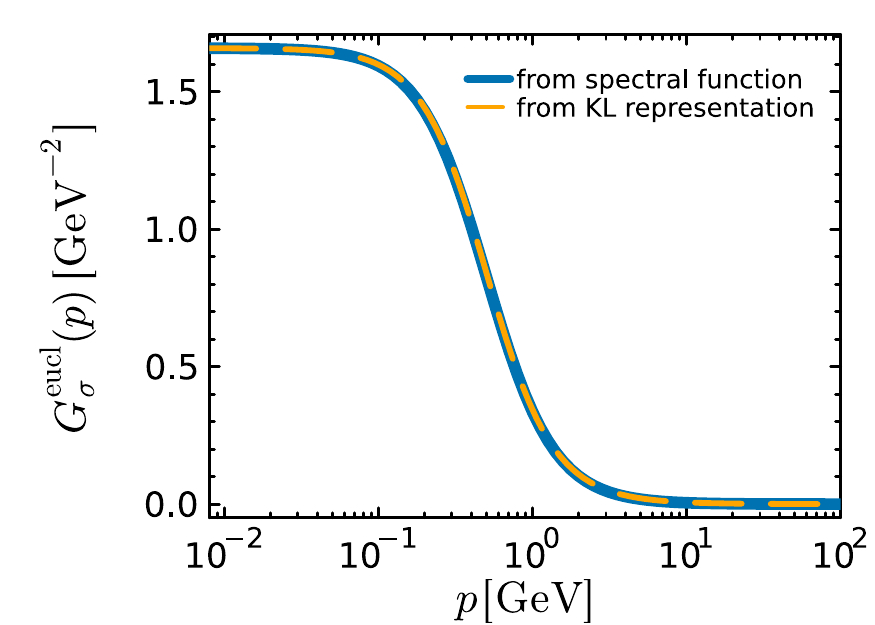}
			\caption{Euclidean $\sigma$-propagator. \hspace*{\fill}}
			\label{fig:KL-ExistenceSigma} 
		\end{subfigure}\hfill
		\begin{subfigure}[t]{0.45\linewidth}
			\includegraphics[width=\textwidth]{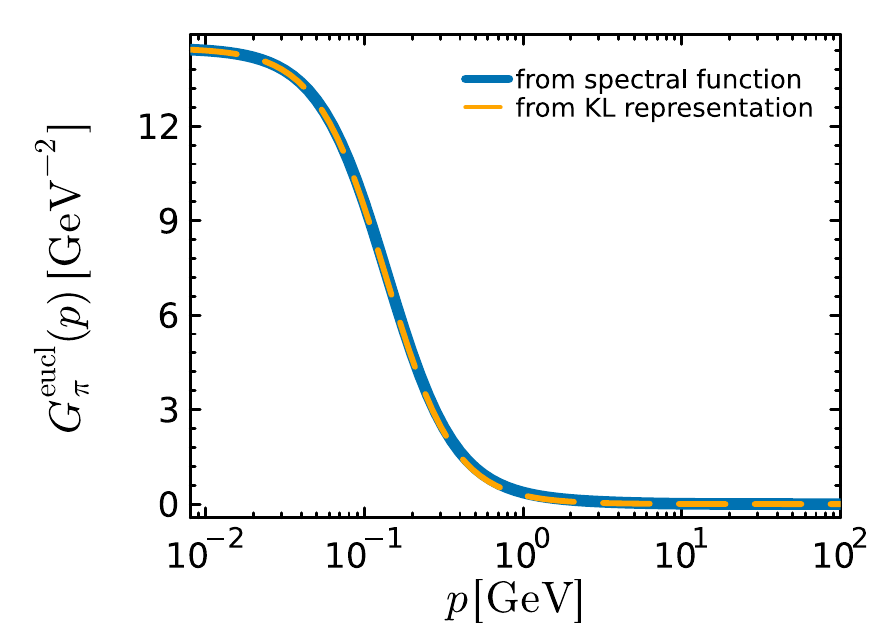}
			\caption{Euclidean pion-propagator. \hspace*{\fill}}
			\label{fig:KL-ExistencePions}
		\end{subfigure}
		\caption{Euclidean meson propagators in comparison to the Euclidean propagator obtained from the KL-representation \labelcref{eq:Meson-KL-Vacuum}.\hspace*{\fill} } 
		\label{fig:KL-ExistenceMesons}
	\end{minipage}
\end{figure*}
%

\section{Existence of the Källén-Lehmann representation for quarks and mesonic composites and RG-consistency}
\label{app:KL-Existence}

\begin{figure*}
	\centering
	\begin{minipage}{0.99\linewidth}
		\centering
		\begin{subfigure}[t]{0.45\linewidth}
			\includegraphics[width=\textwidth]{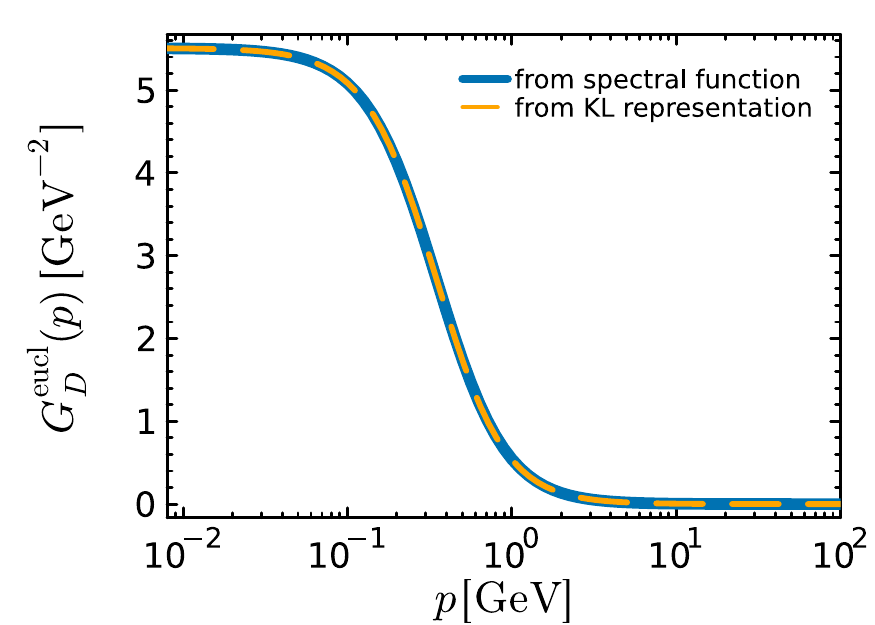}
			\caption{Dirac part \hspace*{\fill}}
			\label{fig:KL-ExistenceDiracQuark} 
		\end{subfigure}\hfill
		\begin{subfigure}[t]{0.45\linewidth}
			\includegraphics[width=\textwidth]{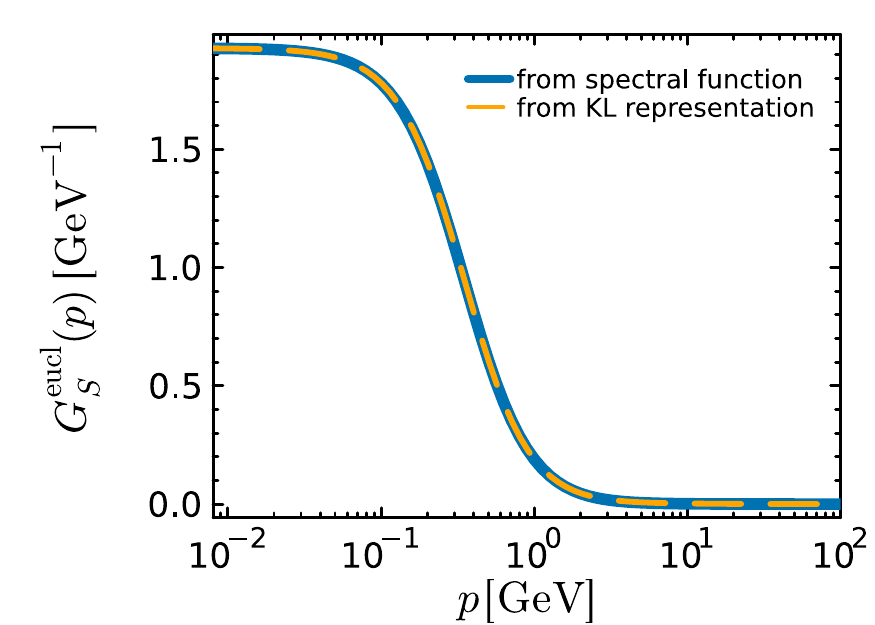}
			\caption{Scalar part \hspace*{\fill}}
			\label{fig:KL-ExistenceScalarQuark} 
		\end{subfigure}
		\caption{Dirac and scalar parts of the Euclidean quark propagator in comparison to that obtained from the KL-representation \labelcref{eq:Quark-KL}. \hspace*{\fill} } 
		\label{fig:KL-ExistenceQuarks}
	\end{minipage}
\end{figure*}

In this Appendix, we assess the existence of the Källén-Lehmann (KL) spectral representation for the quarks and the mesonic composites numerically. If the KL-representation does not hold, the spectral relation \labelcref{eq:Meson-KL-Vacuum} picks up further contributions from singularities in the complex plane, 
\begin{align}
	G_{\phi}(p_0,\boldsymbol{p}) = \int_{0_-}^\infty\frac{\mathrm{d}\lambda}{\pi}\, \frac{\lambda\ \rho_{\phi}(\lambda)}{\lambda^2+p^2}+ \sum_{j \in \text{poles}} \frac{R_j}{p_0^2 + M_j^2}\,.
	\label{eq:GenSpecRep} 
\end{align}
see e.g.~\cite{Cyrol:2018xeq}. In \labelcref{eq:GenSpecRep}, we have omitted possible contributions from complex-conjugate cuts for brevity. We test for additional pole contributions by setting $R_j=0$ and comparing Euclidean propagators obtained in two ways: directly from the spectral flow at Euclidean momenta and from the spectral representation. The two procedures use the absence of the pole terms at different stages of the computation and would disagree if sizeable residues $R_j\neq0$ were present.

A further check is given by computing the Euclidean flow directly and feeding it back. While this check typically provides larger deviations from the spectral result, it requires a slightly different computational setup, and we refrain from adding this. 

We emphasize that this and similar numerical checks can only show the \textit{compatibility} of the KL-representation within the numerical error bounds. For too small residues $R_j\neq 0$ their contributions are within the computational error estimate. Then, however, the violation of the KL-representation is of no practical relevance. 

In \Cref{fig:KL-ExistenceMesons} we show the Euclidean $\sigma$- and pion propagators as well as the Euclidean propagator obtained from the KL-representations \labelcref{eq:Meson-KL}. In \Cref{fig:KL-ExistenceQuarks} we show the scalar and Dirac parts of the Euclidean quark propagator at vanishing cutoff scale, $k=0$ as well as the Euclidean propagator obtained from the KL-representations \labelcref{eq:Quark-KL}. For both quarks and mesons, the KL-representations agree within the numerical accuracy with the Euclidean propagators. Accordingly, the results are compatible with the existence of a Källén-Lehmann spectral representation for quarks and mesonic composites, which serves as a consistency check of our computation.

A second consistency check is the numerical fulfilment of the spectral sum rules~\labelcref{eq:SpecSumRule}. They are fixed by the asymptotic normalisation of $Z_{\Phi^i}(p)$, which in our setup was set to 1. The integrated spectral weights have to match the inverse of this value. The deviation from unity therefore provides a lower estimate of the numerical error.

We compare the sum rules in three computational setups: (1) feeding back only the pole with running $Z_{\Phi^i}$, i.e. neglecting the feedback from the spectral tails; (2) including the spectral tails with a classical initial condition; and (3) the full computation, including spectral tails and the QCD-assisted initial condition. The results are listed in \Cref{tab:SumRule}. 

\begin{table}
	\small
	\renewcommand{\arraystretch}{1.2}
	\setlength{\tabcolsep}{4pt}
	\begin{tabular}{l|c|c|c}
		 & $\int_\lambda\, \lambda\rho_\pi$ & $\int_\lambda\, \lambda\rho_\sigma$ & $\int_\lambda\, \rho_q^D$\\[1ex]
		\hline
		only pole, classical IC & 1.000 & 0.9991 & 0.995 \\
		full $\rho$, classical IC & 1.002 & 1.113 & 1.05 \\
		full $\rho$, QCD-assisted & 1.003 & 1.118 & 1.06 \\
	\end{tabular}
	\caption{Sum rule of the spectral functions in different computation schemes. The deviation from unity provides a lower bound on the numerical error.} 
	\label{tab:SumRule}
\end{table}

The pole-only computation has negligible numerical error. The deviations become substantially larger once higher-order scattering contributions are included. These contributions involve higher-dimensional spectral integrals and are therefore numerically more demanding, leading to deviations of up to 12\%. This is most pronounced for the sigma spectral function, where the $\sigma\to2\sigma$ contribution contains high-dimensional integrals over sharply peaked spectral functions. The QCD-assisted initial condition does not significantly affect this estimate.

This check therefore provides a lower bound on the numerical error of our computation.

\section{Setup details}
\label{app:AdditionalSetup}

In this section we give additional details on the setup of the computation and give the derivations of the formulas used in~\Cref{sec:Setup}.

\subsection{Details on flow and renormalisation of 2- and 4-point functions}

\subsubsection{Computation of the momentum-dependent four-point function}
\label{sec:AdditionalFourPointFunction}
In~\Cref{sec:Setup} we presented the use of the inhomogeneous Bethe-Salpeter equation with a classical scattering kernel to compute the momentum dependence of the four-point function in the $s$-channel, see~\Cref{fig:resummation}.

The pions carry the adjoint representation of the unbroken part of the chiral group, reflected by an index running from 1 to $N_f^2-1$, i.e. 1 to 3 in the case of 2 flavour QCD, giving rise to the quark-meson model. Consequently, the four-pion vertex consists of a sum of several tensor structures with different prefactors. However, in this computation we are only interested in the part that contributes when the vertex is inserted in the tadpole diagram. There, the two external legs as well as the two connected via the regulator line get contracted with a $\delta_{ab}$, making the only non-vanishing contribution proportional to
\begin{align}
	\Gamma^{(4\pi)}_{abcd}\propto\delta_{ad}\delta_{bc}\,,
\end{align}
where we abbreviated
\begin{align}
	\Gamma^{(4\pi)}_{abcd}=\frac{\delta}{\delta\pi^d}\frac{\delta}{\delta\pi^c}\frac{\delta}{\delta\pi^b}\frac{\delta}{\delta\pi^a}\Gamma[\Phi]\,.
\end{align}
The four-pion vertex satisfies the inhomogeneous Bethe-Salpeter equation, which, in the 2PI formalism at leading order in the $1/N$ expansion, reads
\begin{align}
	\Gamma^{(4\pi)}_{abba}(p) = S^{(4\pi)}_{abba} - &\frac{1}{2}\delta_{ab} \sum_c \int\frac{\mathrm{d}^4 q}{(2\pi)^4} S^{(4\pi)}_{aacc} \nonumber\\
	& \times G_{\pi^c}(p+q) \Gamma^{(4\pi)}_{ccaa}(p) G_{\pi^c}(q)\,.
\label{eq:4piVertex}
\end{align}
For a graphical representation, see~\Cref{fig:resummation}. Parts with $a\neq b$ only give momentum-independent contributions and are therefore readily absorbed by the counter term. The only non-trivial contribution comes from the $a=b$ part, given as
\begin{align}
	\Gamma^{(4\pi)}_{aaaa}(p) = \frac{\lambda_{4\pi}}{1+\frac{N_f^2-1}{2}\lambda_{4\pi}D_{\text{fish}}(p)} \,,
\end{align}
with the pion fish diagram, renormalised at $p=0$,
\begin{align}
	D_{\text{fish}}(p) = \int\frac{\mathrm{d}^4q}{(2\pi)^4}\Big\{G_\pi(p+q)G_\pi(q)-G_\pi(q)G_\pi(q)\Big\}\,.
\end{align}
To match this expression at vanishing momentum to the corresponding derivative of the effective potential, we set
\begin{align}
	\lambda_{4\pi} = V''(\rho_0)\,.
\end{align}
If the pion propagator admits a spectral representation, so does the s-channel of the four-pion vertex,
\begin{align}
	\Gamma^{(4\pi)}(p) = \int_0^\infty\frac{\mathrm{d}\lambda\ \lambda}{\pi}\frac{\rho_4(\lambda)}{\lambda^2+p^2}\,,
\end{align}
with the spectral function defined analogously to the propagator spectral function,
\begin{align}
	\rho_4(\omega) = 2\text{Im}\Gamma^{(4\pi)}(p\to -i(\omega +i\varepsilon))\,.
\end{align} 
Inserting this representation into the expression for the tadpole diagram, we get an expression for it in terms of integrals over the spectral function of the vertex ($\rho_4$) and the regulator line ($\rho_{\pi^2}$),
\begin{align}
	D_{\text{tad}}(p)&=\int\frac{\mathrm{d}^4q}{(2\pi)^4} \Gamma^{(4\pi)}(p+q)G_\pi^2(q) \nonumber\\
	&= \int\frac{\mathrm{d}\lambda_1 \mathrm{d}\lambda_2}{\pi^2}\ \lambda_1\lambda_2 \rho_4(\lambda_1)\rho_{\pi^2}(\lambda_2) I_{\text{pol}}(p,\lambda_1,\lambda_2)\,.
	\label{eq:tadpoleSpectral}
\end{align}
$I_{\text{pol}}$ is the analytic integration kernel resulting from the integration over the loop momentum, see~\Cref{app:kernels} for explicit expressions. The spectral function of the regulator line is given in~\Cref{app:regulator_lines}. Notably, this expression is structurally equivalent to the expression of the polarisation diagrams.

\subsubsection{Renormalisation of the 2-point functions}
\label{sec:Renormalisation}

In the fRG-framework, the regulator suppresses infrared modes. When choosing a sufficiently fast decaying regulator (like exponential regulators) or even ones with compact support (like the Litim-regulator), this suppression does not extend to the UV modes. Consequently, in these cases changing the regulator scale mainly affects the momentum modes around the scale $k$, while the UV remains largely unaffected.

However, the aforementioned decaying regulators break either causality (through the introduction of non-analyticities in the complex $p^0$ plane) or Lorentz-invariance (if they are chosen to only depend on the spatial momentum $\vec{p}$). As these two features of the theory in vacuum are essential for the use of spectral functions, decaying regulators are not suited for this work, see~\cite{Roth:2021nrd,Braun:2022mgx}.

In contrast, the Callan-Symanzik (CS) regulator,
\begin{align}
	R_{CS} = Z_{\phi}k^2\,,
\end{align}
is momentum independent, and thus trivially preserves Lorentz-invariance and causality. Since it does not decay in the UV, it does not lead to a priori finite flows if naively inserted into the flow equations. 

Nevertheless, the insertion of a CS-regulator lowers the degree of divergence of all flows of operators by two, compared to other diagrammatic approaches like the Dyson-Schwinger equations. Thus, the flow of marginal couplings, like the on-shell wave function $Z_{\pi/\sigma}$ or the four-meson vertices, is finite, and only the initial condition requires regularisation.

In contrast, for some relevant operators, like the masses of both mesons and quarks, lowering the degree of divergence is not enough to obtain finite flows. For example, the flow of the mesonic mass parameter is logarithmically divergent, due to the contributions of the constant part of the tadpole and the fermionic polarisation diagram. These UV divergences are subtracted by the introduction of a flowing counter-term action. For details on the precise nature of this action, see~\cite{Braun:2022mgx}. Here, it suffices to say that this counterterm is not an ad-hoc construction but arises naturally, when the CS-equation is defined from a limit of finite flow-equations. 

The coefficients of the counter-term action are fixed by \textit{flowing} renormalisation conditions. These allow one to not only fix the value of the respective term at $k=0$, but their whole trajectory during the flow, omitting any fine-tuning of the initial condition.

In practice, this means, that the flow of the two-point functions of both mesons and quarks are fixed up to a real constant $\partial_tS_{\text{ct}}^{(2)}$. This constant can be set such, that the scale dependence of the meson pole masses are fixed to a tree-level trajectory, which ends on the physical points at vanishing cutoff scale
\begin{align}\nonumber
	0 & = \left[\Re\Gamma^{(2\pi)}_k(p^2) + R_{\phi} + \partial_tS_{\text{ct},\pi}^{(2)}\right]\bigg|_{p^2=-m_{\pi,\text{\tiny{pole}}}^2}\,,\\[1ex]
	0 & = \left[\Re\Gamma^{(2\sigma)}_k(p^2) + R_{\phi} + \partial_tS_{\text{ct},\sigma}^{(2)}\right]\bigg|_{p^2=-m_{\sigma,\text{\tiny{pole}}}^2}\,,
	\label{eq:renormalisation_condition_mesons}
\end{align}
with the pole masses
\begin{align}\nonumber
	m_{\pi,\text{\tiny{pole}}}^2(k) & = m_{\pi}^2+k^2\,,\\[1ex]
	m_{\sigma,\text{\tiny{pole}}}^2(k) & = m_{\sigma}^2+k^2\,.
	\label{eq:meson_pole_mass_trajectory}
\end{align}
The parameters $m_{\pi}$ and $m_{\sigma}$, i.e. the pole masses at $k=0$ are the physical masses of the mesons and are input parameters. Since the pions correspond to real existing particles, their mass can be chosen to be their measured mass. Here, we set $m_\pi = 138\ \mathrm{MeV}$, to be compatible with previous isospin-symmetric QCD computations. For the $\sigma$, we use a Breit-Wigner mass of $m_\sigma = 500\ \mathrm{MeV}$.

For the quarks, there is a slightly different renormalisation condition. Instead of the pole mass, we instead fix the trajectory of the constituent quark mass, i.e. the value of the quark mass function $M$ at vanishing momentum,
\begin{align}
	M(p^2=0)+ Z_q^{-1} R_q = m_{\mathrm{const}} + k\,.
	\label{eq:renormalisation_condition_quarks}
\end{align}
Similarly to the mesonic case, the constituent mass $m_{\mathrm{const}}$ is an input parameter. To be consistent with previous works, we choose a mass of $\SI{350}{\MeV}$.

\begin{figure*}
	\centering
	\begin{minipage}{0.99\linewidth}
		\centering
		\begin{subfigure}[t]{0.48\linewidth}
			\includegraphics[width=\textwidth]{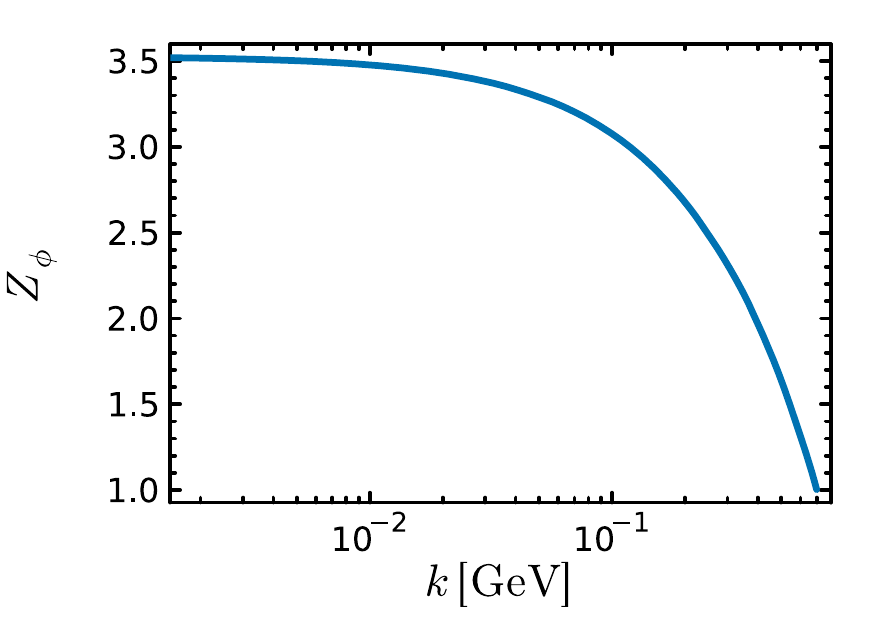}
			\caption{Mesonic on-shell wave function $Z_\phi$. \hspace*{\fill}}
			\label{fig:Z_phi}
		\end{subfigure}\hfill
		\begin{subfigure}[t]{0.48\linewidth}
			\includegraphics[width=\textwidth]{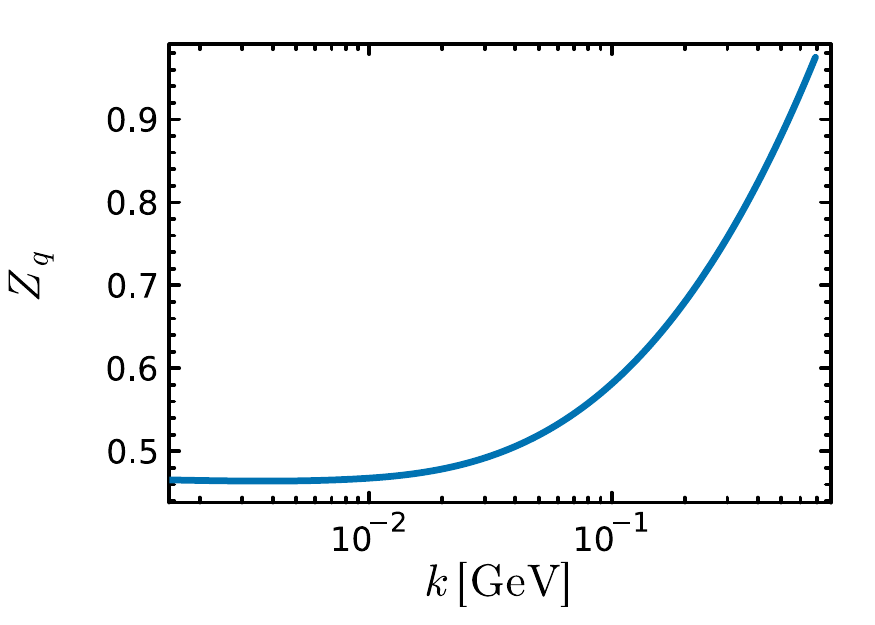}
			\caption{Fermionic on-shell wave function $Z_q$. \hspace*{\fill}}
			\label{fig:Z_q}
		\end{subfigure}
		\caption{Scale-dependence of the on-shell wave functions $Z_\phi$ and $Z_q$, defined as the residuals of the propagator on the mass pole.
		\hspace*{\fill} } 
		\label{fig:Zs}
	\end{minipage}
\end{figure*}

For completeness, we state again that the wave function renormalisations $Z_{\Phi_i}(p)$ are fixed to unity for $p^2 \rightarrow \infty$ for all scales $k$. The predicted running of the on-shell wave functions is shown in~\Cref{fig:Zs}.

In addition to the masses and wave functions, there is another parameter whose flow we can fix. That parameter is the explicit symmetry breaking $c_\sigma$ coming from the current quark mass. The associated operator does not explicitly enter the flow, since it is linear in the field, and the flow does only depend on the second (and higher) derivatives of the effective action. What $c_\sigma$ determines instead, is the expansion point $\rho_0$, i.e. the position of the physical minimum and solution of the equation of motion
\begin{align}
	\left[V'(\rho)-\frac{c_\sigma}{\sqrt{2\rho}}\right]\bigg|_{\rho=\rho_0}=0\,.
\end{align}
Letting $c_\sigma$ flow on a fixed trajectory allows us to fix the position of the physical minimum $\rho_0$, or equivalently the expectation value $\sigma_0=\sqrt{2\rho_0}$. In our approximation, this expectation value coincides with the pion decay constant $f_\pi=\SI{93}{\MeV}$.

\subsubsection{Emergent composites and dynamical hadronisation}
\label{sec:setupDynhad}

As described in~\Cref{sec:EC}, we capture the dynamics of the resonant scalar-pseudoscalar channel in the four-quark vertex via the introduction of a mesonic field $\phi$. In this section, we explain more technical details on this way of dynamically hadronising, specifically its concrete implementation.

Through this mesonic field, the full four-fermion vertex can be reparametrised as a sum of a meson exchange and a residual part, see~\labelcref{eq:ResonantResidualSplit}. By requiring the residual part to vanish in the $t$-channel, i.e.~\labelcref{eq:tChannelConstraint}, the mesonic field carries the full information of the resonance.

In order for this constraint to be fulfilled at every scale, the mesonic field $\phi$ is promoted to a field-dependent operator,
\begin{align}
	\phi\ \to\ \phi_k\,.
\end{align}
Since the fields now carry an explicit $k$-dependence as well, this modifies the original Wetterich equation, leading to the \textit{generalised} flow equation, see~\cite{Pawlowski:2005xe}. While this transformation can in principle be done in a fully field- and momentum-dependent way, in this work we limit it to a truncated, momentum independent transformation,
\begin{align}
	\dot{\phi}_a=\dot{A}_k\, \bar{q}\tau_a q\,,
\end{align}
with $\tau$ defined in~\labelcref{eq:phi-rho}, and the scalar hadronisation function $\dot{A}_k$. This choice of field transformation leads to a modification of the flows of the Yukawa and four-fermion interactions. Schematically, the modified flows of the Yukawa interactions read
\begin{align}
	\frac12\partial_t h_\phi + \Gamma^{(2)}_{k,\phi\phi}\dot{A}_k & = \text{Flow}_{\Gamma^{(3)}_{k,\Bar{q}q\phi}}\,, \quad \phi=\sigma,\pi\,,
\end{align}
where $\text{Flow}_{\Gamma^{(n)}}$ consists of the diagrams obtained by taking the corresponding derivatives of the RHS of the Wetterich equation. As the momentum dependence of the Yukawa coupling can be absorbed by the mesonic propagator, we restrict ourselves to the flow at vanishing momentum.

The modified flow of the (residual) four-fermion vertex reads
\begin{align}
	\partial_t\lambda_{q,\textrm{res}}+ \frac{h_\phi}{2} \dot{A}_k & = \text{Flow}_{\Gamma^{(4)}_{k,\Bar{q}q\Bar{q}q}}\,,
	\label{eq:flow_4fermi_dynhad}
\end{align}
where the diagrams on the right-hand side have been projected onto the scalar-pseudoscalar channel. Fixing the constraint $\partial_t\lambda_{q,\textrm{res}}=0$ in the $t$-channel uniquely fixes $\dot{A}_k$ as a function of the Yukawa coupling and diagrams,
\begin{align}
	\dot{A}_k = \frac{2\,\textrm{Flow}_{\Gamma^{(4)}_{\Bar{q}q\Bar{q}q}}}{h_\phi}\,.
\end{align}
As this hadronisation function is part of the flow of the Yukawa coupling, the absorption of the resonance dynamics in the meson exchange becomes evident.

\subsubsection{Iteratively solving the flow equation} 

Due to the presence of non-analyticities of the two-point functions (the onsets of the imaginary parts), explicit solving methods, like Runge-Kutta methods, are unstable and introduce numerical errors. Since the position of these non-analyticities is known, they can be conveniently circumvented using fixed-point iteration. In this section, we outline this method of solving the spectral flow equation, which was developed in~\cite{Pawlowski:2025etp}. Schematically, the equation we solve is of the form 
\begin{align} 
	\partial_t\Gamma^{(2)}(p,k) = \text{Flow}\left[\{\Gamma^{(n)}\}, \rho\right](p,k)\,, 
\end{align} 
where $n=3,4$ and $\Gamma^{(2)}$ and $\rho$ are the two-point and spectral functions of the particles, respectively. The term $\text{Flow}[\{\Gamma^{(n)}\}, \rho](p,k)$ on the right-hand side of the flow equation comprises the diagrams. The latter depends on the spectral functions as well as the three- and four-point functions. The non-analyticities of the flow equation are fully determined by the form of the spectral functions and the diagram topologies. Thus, if these spectral functions are known, the exact position of the branch-points are as well. To use this advantage, we solve the flow iteratively, i.e. 
\begin{align} 
	\partial_t\Gamma_{i+1}^{(2)}(p,k) = \text{Flow}\left[\{\Gamma_i^{(n)}\},\rho_i\right](p,k)\,. 
\end{align} 
Consequently, the flow equation of the iteration $i+1$ only depends on the results of the previous iteration. This turns the flow equations into a system of integral equations,
\begin{align}
	\Gamma_{i+1}^{(2)}(p,k) = \int_\Lambda^k\frac{\mathrm{d} k}{k}\,\text{Flow}\left[\{\Gamma_i^{(n)}\},\rho_i\right](p,k) + \Gamma^{(2)}(p,\Lambda)\,. 
\end{align} 
$\Lambda$ is the boundary of the $k$-integration and $\Gamma^{(2)}(p,\Lambda)$ the initial condition which does not vary with $i$. To close this procedure, we have to choose an initial set of spectral functions $\rho_0$. For that, we use free spectral functions, containing only the one-particle delta pole. The next iteration of $\Gamma^{(2)}$ is computed on a $p\times k$ grid and interpolated. This gives the spectral functions of the next iteration, which are then plugged back into the flow equation, computing the following iteration. The non-analyticities can be taken explicitly into account by including them in the boundaries of the spectral integrals. If this scheme converges, the limit is a solution of the flow equation. To hasten the convergence, we update only the $Z$s, i.e. the prefactor of the one-particle poles in the first few iterations, while neglecting any contribution from the scattering tails. Only after this approximation is converged, we begin the fully self-consistent iteration. We found that this speeds up the numerical computation considerably, since the contributions of the one-particle poles are computationally cheap and dominate in a large range of the flow.

\subsection{Flow of the full potential}
\label{subsec:Potential_flow}

\begin{figure*}
	\centering
	\begin{minipage}{0.99\linewidth}
		\centering
		\begin{subfigure}[t]{0.45\linewidth}
			\includegraphics[width=\textwidth]{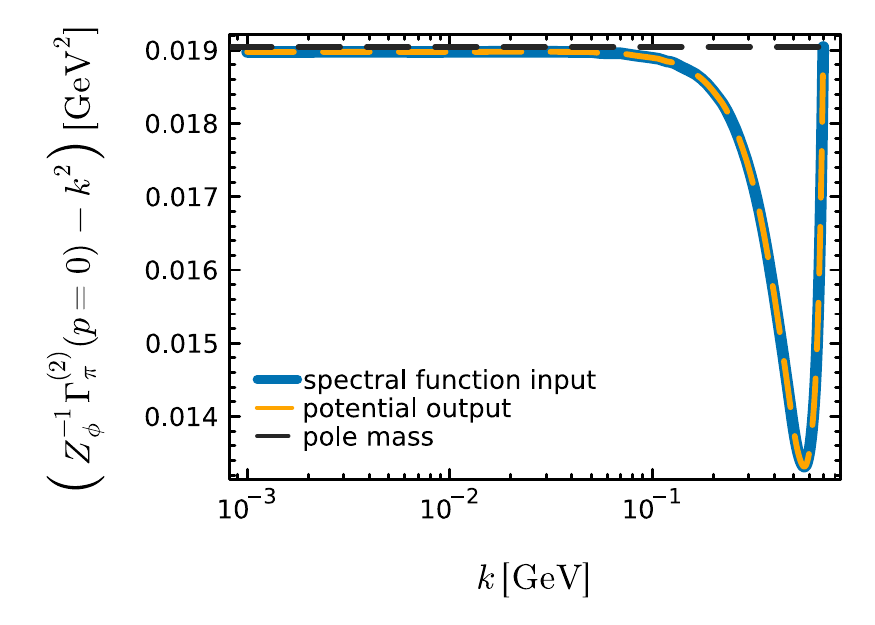}
			\caption{Pion curvature mass. \hspace*{\fill}}
		\end{subfigure}\hfill
		\begin{subfigure}[t]{0.45\linewidth}
			\includegraphics[width=\textwidth]{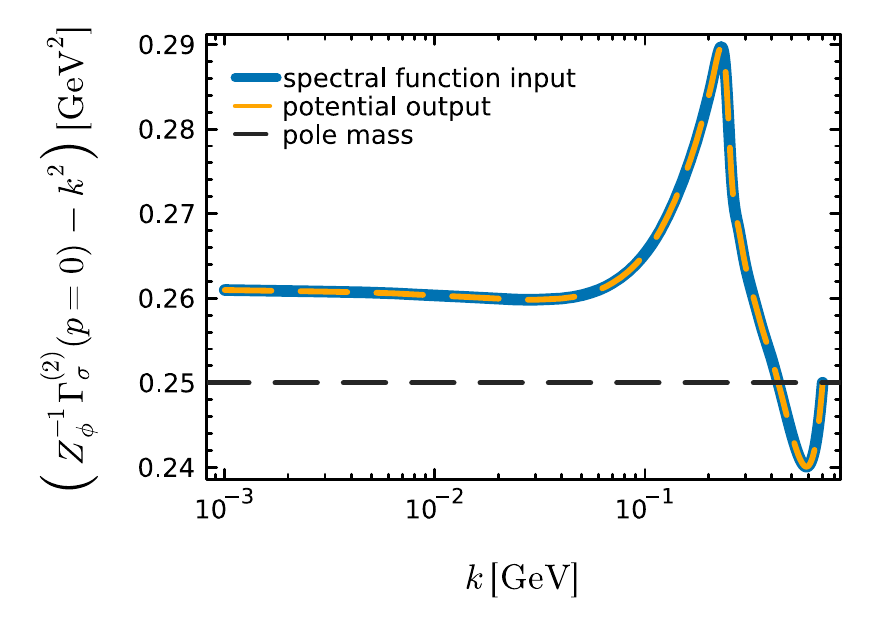}
			\caption{Sigma curvature mass. \hspace*{\fill}}
		\end{subfigure}
		\caption{Scale dependence of two mesonic curvature masses, i.e. the RG-invariant inverse propagator at $p=0$ with regulator subtracted. These values serve as renormalisation conditions for the effective potential, see~\labelcref{eq:PotentialFromSpecFunc}. We show both the input computed from the spectral functions and the output from the effective potential. The excellent agreement between the two serves as a numerical cross-check for the renormalisation procedure of the effective potential. Notably, the kink in the sigma curvature mass at $k\approx0.24\ \mathrm{GeV}$ coincides with the scale, at which the sigma becomes unstable, see also~\Cref{fig:PotentialDerivatives}. The renormalisation condition~\labelcref{eq:RenormCondMesons} results in this non-smooth behaviour. \hspace*{\fill} } 
		\label{fig:PropagatorsZeroMom}
	\end{minipage}
\end{figure*}

We now give additional details on the flow of the effective potential, namely the explicit expression for the flow equation and the regulator used in this computation.

\subsubsection{Flow equation of the full effective potential}

To compute the flow of the effective potential, we use the form of the flow defined in~\labelcref{eq:FinalRen-flow}. In the approximation detailed in~\labelcref{eq:PotentialPropTrunc}, the flow reads, dropping the prime from the mesonic field,
\begin{align}\nonumber
	\partial_t V_k(\rho) &= \frac{1}{2}\,\text{Loop}_\sigma(\rho) + \frac{(N_f^2-1)}{2}\,\text{Loop}_\pi(\rho)\\
	&\quad - N_fN_c \,\text{Loop}_q(\rho) - \dot{\phi}\,\partial_\phi V(\rho) + \partial_tS_{\text{ct}}\,,
	\label{eq:PotentialFlow}
\end{align}
with $N_f=2$ flavours and $N_c=3$ colours. For a diagrammatic representation, see~\Cref{fig:FlowofGamma}. The three loop contributions are given by
\begin{subequations}
	\begin{align}
		\text{Loop}_\phi(\rho)&=\int\frac{\dd^4p}{(2\pi)^4}\ \frac{\dot{R}_\phi}{\Gamma^{(2)}_{\phi,\text{dyn}}(p)+m_\phi^2(\rho)+R_\phi}
		\label{eq:PotentialFlowLoops_phi}
	\end{align}
	\begin{align}
		\text{Loop}_q(\rho)&=\Tr\int\frac{\dd^4p}{(2\pi)^4}\ \frac{\dot{R}_q}{Z_q(p)(i\slashed{p}+M(p,\rho))}\nonumber\\
		&=\int\frac{\dd^4p}{(2\pi)^4}\ \frac{1}{Z_q(p)}\frac{M(p,\rho)\,\dot{R}_q}{p^2+M(p,\rho)^2}\,,
		\label{eq:PotentialFlowLoops_q}
	\end{align}
	\label{eq:PotentialFlowLoops}
\end{subequations}
with $\phi=\pi,\sigma$. Notably, in the present setup with a shifted mesonic field, the quark loop no longer contains the explicit regulator, only its derivative.

All three diagrams are divergent and are finite only after renormalisation via the counter term $\partial_tS_{\text{ct}}$. In the following, we discuss how the renormalisation conditions \labelcref{eq:renormalisation_condition_mesons,eq:renormalisation_condition_quarks} are implemented in the flow of the effective potential.
%

\subsubsection{Regularisation and renormalisation}
\label{sec:PotentialRegularisation}

With a regulator that does not decay in the UV, the flow of the effective potential~\labelcref{eq:PotentialFlow} and its derivatives diverge in $d=4$. Thus, it has to be regularised by subtracting appropriate counter terms. As outlined in \cite{Braun:2022mgx}, in an LPA truncation the counter terms for the mesonic loops are proportional to the mesonic masses. This dependence also appears in the truncation used in this work, see~\labelcref{eq:PotentialPropTrunc}.

\paragraph{Mesonic loop}

We start with a discussion of the regularisation of the mesonic loops. As they are quadratically divergent, they show a divergence proportional to the mesonic mass. In order to render these diagrams finite, a term proportional to the (field dependent) mass has to be subtracted. This results in
\begin{widetext}
	\begin{align}\nonumber
		\text{Loop}_\phi^{\text{finite}}(\rho)=\int\frac{\dd^4p}{(2\pi)^4}\ \bigg[&\frac{\dot{R}_\phi}{\Gamma^{(2)}_{\phi,\text{dyn}}(p)+m_\phi^2(\rho)+R_\phi}\\
		&-\frac{\dot{R}_\phi}{\Gamma^{(2)}_{\phi,\text{dyn}}(p)+R_\phi}+\frac{\dot{R}_\phi m_\phi^2(\rho)}{(\Gamma^{(2)}_{\phi,\text{dyn}}(p)+m_{\phi,0}^2+R_\phi)(\Gamma^{(2)}_{\phi,\text{dyn}}(p)+R_\phi)}\bigg]\,,
	\end{align}
\end{widetext}
with the renormalisation scale $m_{\phi,0}^2$. The value of this renormalisation scale has to approach the sigma mass from above in order to avoid negative diffusion in the equation of the effective potential which would render the simulation unstable.

\paragraph{Quark loop}

A similar procedure is used to render the quark loop finite. As outlined in~\labelcref{eq:PotentialFlowLoops}, the expression of the quark loop has a very similar form as the mesonic loop. In particular, the divergence structure is the same. Thus, they require a similar subtraction scheme to render them finite.

The fermionic mass is proportional to the field, and since it is already subject to renormalisation, it generates counter terms proportional to $\phi$, $\rho$, and $\rho^2$ which are fixed by the renormalisation of the 2-point functions, see~\labelcref{sec:Renormalisation}, and therefore dictate the renormalisation of the fermionic contributions to the effective potential.

\begin{figure*}[t]
	\centering
	\begin{minipage}[t]{0.48\linewidth}
		\centering
		\includegraphics[width=0.95\textwidth]{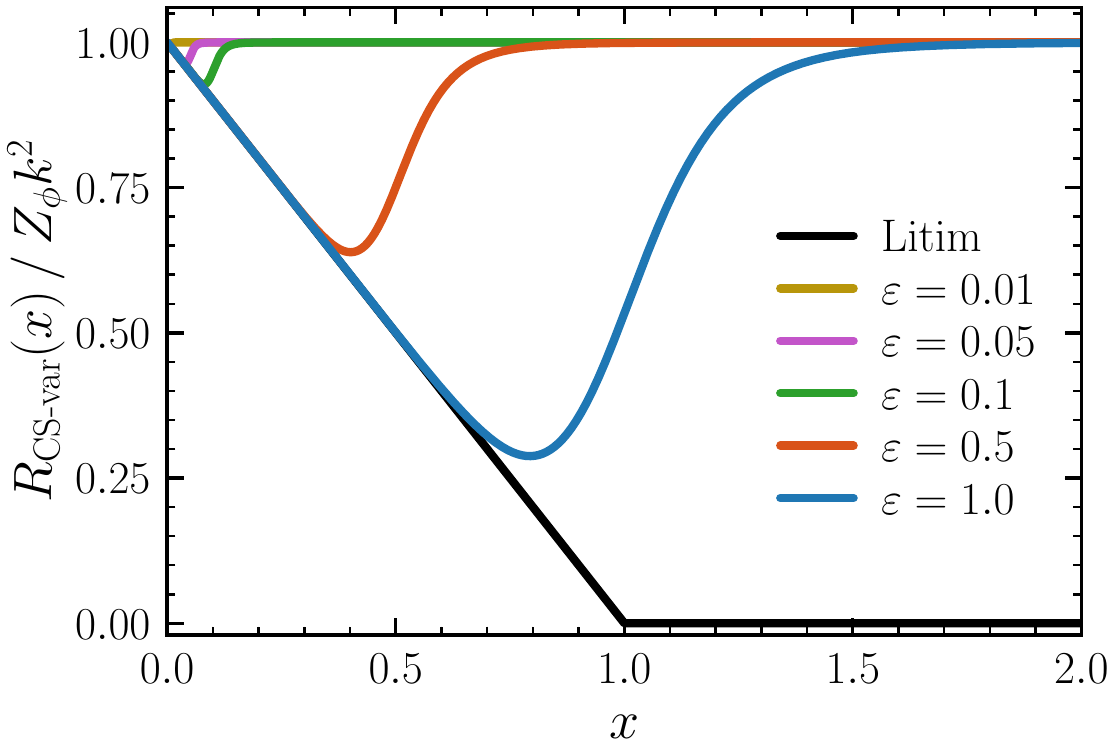}
		\caption{Shape function of the CS-variant regulator \Cref{eq:CSvar} with $n=8$ and $m=10$ for different values of $\varepsilon$. For comparison, we also show the Litim regulator.\hspace*{\fill}}
		\label{fig:CSvariant_regulator}
	\end{minipage}%
	\hspace{0.03\linewidth}%
	\begin{minipage}[t]{0.48\linewidth}
		\centering
		\includegraphics[width=0.95\textwidth]{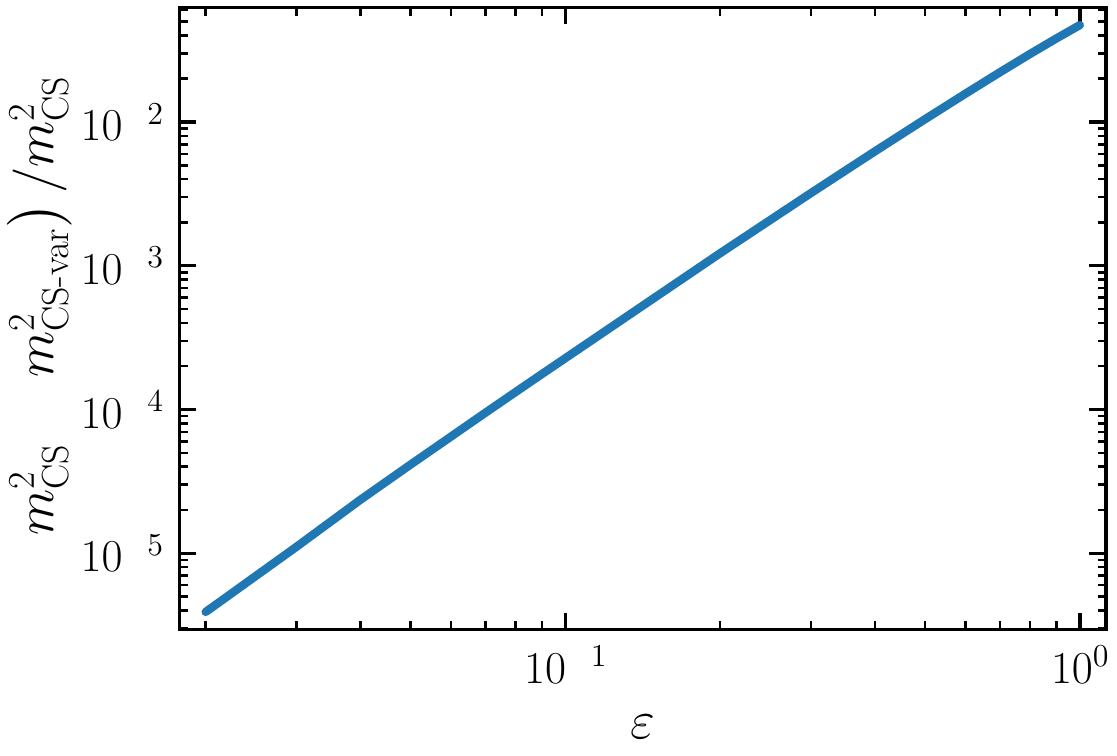}
		\caption{Difference between the CS and CS-variant regulator as a function of the parameter $\varepsilon$. To quantify this difference, the flow of a symmetric $O(4)$ model was computed using varying $\varepsilon$. All flows were performed with the same initial conditions. We compare the curvature of the potential at $\rho=0$ and $k=0$ ($m^2$). The dependence yields a power law in $\varepsilon$. Already at $\varepsilon=0.1$, there is a relative difference of $\sim 10^{-4}$. We conclude that we can extrapolate computations with the CS-variant regulator to $\varepsilon=0$, the limit of the Callan-Symanzik regulator. \hspace*{\fill}}
		\label{fig:CSvariant_effect}
	\end{minipage}
\end{figure*}
%

\paragraph{Renormalisation conditions and full expression}

There are two ways to compute the first two potential derivatives on the minimum, which must coincide: (1) the $p\to 0$ of the two-point functions~\labelcref{eq:PotentialFromSpecFunc}, and (2) the derivatives of the effective potential computed via~\labelcref{eq:PotentialFlow}. Thus, we use~\labelcref{eq:PotentialFromSpecFunc} as renormalisation condition. Since we can additionally introduce a counter term for the explicit symmetry-breaking term $c_\sigma\sigma$, which determines the position of the physical minimum $\rho_0$, we can choose it to be independent of the RG scale $k$. Hence, we can set the field expectation value to a constant value
\begin{align}
	\rho_0=\frac{f_\pi^2}{2}\,,
\end{align}
with the pion decay constant $f_\pi=\SI{93}{\MeV}$, which in our setup corresponds to the expectation value of the sigma.

Thus, the full, renormalised flow equation for the effective potential reads:
\begin{align}
	\partial_t V_k(\rho) = \text{Flow}^\text{finite}(\rho) + \partial_t\tilde{S}
\end{align}
with $\text{Flow}^\text{finite}(\rho)$ the sum of the finite loops, similar to~\labelcref{eq:PotentialFlow}. The remaining counter term is given by:
\begin{align}\nonumber
	\partial_t\tilde{S} = &- \rho\left(\partial_\rho\text{Flow}^\text{finite}\right)\bigg|_{\rho=\rho_0} - \frac{\rho^2}{2}\left(\partial_\rho^2\text{Flow}^\text{finite}\right)\bigg|_{\rho=\rho_0}
	\\[1ex]
	&+ \partial_tm_{\pi,\text{curv}}^2 \rho + \left(\frac{\partial_tm_{\sigma,\text{curv}}^2-\partial_tm_{\pi,\text{curv}}^2}{2\rho_0}\right)\frac{\rho^2}{2}
	\nonumber\\[1ex]
	&+\partial_t c_\sigma \sigma
\end{align}
While the last term does not directly enter the flow, it enters indirectly, as it allows us to fix the position of the minimum to a constant value, $\partial_t\rho_0 = 0$.

The $k$-dependence of the mesonic propagators at vanishing momentum is shown in \Cref{fig:PropagatorsZeroMom}, where we compare values extracted from the spectral functions against those produced directly by the potential flow. The agreement between the two provides a non-trivial consistency check, demonstrating that the renormalisation scheme can be implemented in a self-consistent manner.

\subsubsection{Choice of regulator}
\label{sec:PotentialRegulator}

To be consistent with the spectral flows and match the scale dependence of the results, both the spectral function and the potential computation require the use of the same regulator. While the CS regulator was designed for the spectral part, it has shortcomings for the flow of the potential. 
A Callan-Symanzik regulator changes the pole of the Wetterich equation at the pion mass $m_\pi^2 = -k^2$, which is associated with the convexity bound, to be an integrable singularity. This makes the convexity restoring property of the flow equation numerically inaccessible, see~\cite{Ihssen:2024miv,Zorbach:2024zjx} for more discussions.
To remedy this problem and increase the divergence of the pole, the order of the singularity in the flow needs to be increased.

To get both numerical access to the convexity restoration and the UV behaviour of the CS cutoff, we use the \textit{CS-variant} regulator~\labelcref{eq:CSvar}. It is given as
\begin{align}
	\frac{R_{\text{CS-var}}(x)}{Z_\phi k^2} &= s_n(x)\cdot(1-b_{m,\varepsilon}(x)) + b_{m,\varepsilon}(x)
	\label{eq:CSvar}
\end{align}
with
\begin{subequations}
	\begin{align}
		s_n(x)&= \exp\left(-\sum_{i=1}^n\frac{x^i}{i}\right)\\[1ex]
		b_{m,\varepsilon}(x) &= \frac{x^m}{x^m+\varepsilon^m}\,.
	\end{align}
\end{subequations}
In~\Cref{fig:CSvariant_regulator} we show the shape of this class of regulators. It depends on a tuning parameter $\varepsilon$ that interpolates smoothly between the momentum-independent CS regulator at $x\gg\varepsilon$, and a polynomial-exponential regulator at $x\ll\varepsilon$. The latter, introduced in~\cite{Ihssen:2024miv}, provides numerical access to the convexity restoration by amplifying the pole in the Wetterich equation to be non-integrable.

Evidently, this procedure introduces a slight inconsistency between the spectral and potential flows. To quantify the effects of the CS-variant regulator in comparison with the standard CS regulator, we study the flow of the effective potential of a 3D $O(4)$ model in the symmetric phase. In this regime, the potential remains convex for all scales $k$, and only the flow of the absolute value of the potential itself requires regularisation. These two facts make this model an ideal testing ground for this type of regulator.

As observable, we choose the curvature of the potential at $\rho=0$ and $k=0$. \Cref{fig:CSvariant_effect} shows the relative deviation of this curvature from the result obtained with the pure CS regulator as a function of $\varepsilon$. Even at $\varepsilon=0.1$, the relative difference is of order $10^{-4}$. This makes the CS-variant regulator a practical numerical tool to approximate the Callan-Symanzik flow, when a direct computation is not numerically stable.

For our computation we use $n=8$, $m=10$ and $\varepsilon=0.3$. Since this convexity restoration is only a property of the mesonic part of the flow, for the quark loop we can use a scalar CS cutoff $R_q=Z_q k$.

\section{Integration kernels}
\label{app:kernels}

The use of spectral representations allows a semi-analytic computation of diagrams containing full propagators and/or vertices in Minkowskian spacetime due to the analytic computation of the loop integrals. While the integrations over the spectral kernels in some cases result in UV-divergent integrals, these divergences can be taken into account using the spectral BPHZ renormalisation scheme, introduced in~\cite{Horak:2020eng}.

The diagrams present in the flow all share the same topology — that of a polarisation diagram. Even though the pion flow contains a tadpole, the inclusion of a spectral function for the four-pion vertex in an s-channel approximation leads to the same type of spectral integral and thus the same topology. Integrals of this type contain the following integration kernels,
\begin{align}
	I^{ij}_{\text{pol}}(p,\lambda_1,\lambda_2)=\int\frac{\mathrm{d}^4 q}{(2\pi)^4}\ \frac{q^i(p+q)^j}{(\lambda_1^2+(p+q)^2)(\lambda_2^2+q^2)}\,.
\end{align}
While the real part of these integrals diverges, their imaginary parts in the retarded limit $p\to -i(\omega + i\epsilon)$ are finite. After performing the integrals, these read
\begin{align}\nonumber
	&\text{Im}I^{00}_{\text{pol}}(\omega,\lambda_1,\lambda_2) = \frac{\zeta}{16\pi^2\omega^2}\left[\text{Im}\text{Atanh}\left(\frac{\lambda_1^2-\lambda_2^2+\omega^2}{\zeta}\right)\right.\\\nonumber
	&\hspace{4cm}\left.-\text{Im}\text{Atanh}\left(\frac{\lambda_1^2-\lambda_2^2-\omega^2}{\zeta}\right)\right]\\\nonumber
	&\text{Im}I^{11}_{\text{pol}}(\omega,\lambda_1,\lambda_2) = \frac{\lambda_1^2+\lambda_2^2-\omega^2}{2}\text{Im}\,I^{00}_{\text{pol}}(\omega,\lambda_1,\lambda_2) \\
	&\text{Im}I^{20}_{\text{pol}}(\omega,\lambda_1,\lambda_2) = \lambda_2^2\ \text{Im}\,I^{00}_{\text{pol}}(\omega,\lambda_1,\lambda_2) \,,
\end{align}
with the cut function
\begin{align}
	\zeta = \sqrt{\lambda_1^4+(\lambda_2^2-\omega^2)^2-2\lambda_1^2(\lambda_2^2+\omega^2)}\,.
\end{align}
Notably, the imaginary part only has support for
\begin{align}
	\lambda_1+\lambda_2\leq \omega\,,
\end{align}
which is also apparent from the cutting rules formulated by Cutkosky \cite{Cutkosky:1960sp}. As a consequence, when computing the imaginary part of a diagram for fixed external momentum $\omega$, the spectral integral has a compact support. Since the spectral functions consist of a continuous scattering tail and a one-particle delta contribution, the imaginary part of a diagram is always finite. 

The real part on the other hand is still divergent as it carries the local divergence of the loop-integral. To make it finite and well-defined it requires a subtraction. This is reflected in the fact that the imaginary part does not decay fast enough in the limit $\omega\to\infty$, and hence the standard Kramers-Kronig relations are not applicable. However, the momentum-independent divergence can be subtracted by computing differences between two momenta. For the two-point function, this gives
\begin{align}
	\Re\,\Gamma^{(2)}(\omega) = &\Re\,\Gamma^{(2)}(\omega_0) \nonumber\\
	 + (\omega^2-\omega_0^2)& \ \mathrm{PV}\int_0^\infty\frac{\mathrm{d}\lambda}{\pi}\,
	\frac{2\lambda\,\Im\,\Gamma^{(2)}(\lambda)}{(\lambda^2-\omega^2)(\lambda^2-\omega_0^2)}\,,
	\label{eq:reducedKK}
\end{align}
with $\omega_0$ the renormalisation point. Evidently, this also generalizes to higher (local) divergencies. This example illustrates that this allows for an efficient implementation of a BPHZ renormalisation scheme with renormalisation conditions discussed in~\Cref{sec:Renormalisation}. Setting $\omega_0^2=m_\phi^2$ and $\Gamma^{(2)}(\omega_0)=0$, the solution of~\labelcref{eq:reducedKK} automatically fulfils the mesonic renormalisation conditions~\labelcref{eq:RenormCondMesons}, and similarly for the fermions.

For more details on the implementation of renormalisation conditions for spectral flows see \cite{Pawlowski:2025etp}.

\section{Spectral functions of regulator lines}
\label{app:regulator_lines}

To reduce the dimensionality of the spectral integrals, we introduce spectral functions for the squared propagator, both mesonic and fermionic. Since the CS-regulator is momentum independent, the momentum dependence of the regulator line in each diagram is given by that of the squared propagator. This allows us to represent the regulator line of a scalar propagator (omitting the regulator derivative) by
\begin{align}\label{eq:Gsq_rep_scalar}
	G^2(p) = \int_\lambda\frac{\rho_{G^2}(\lambda)}{\lambda^2+p^2}\,.
\end{align}
This straightforwardly extends to all kinds of propagators which have a KL-representation, if their square is projected onto the respective tensor structures. We make it explicit for Dirac fermions, in our case the quarks:
\begin{align}\nonumber
	G_q^2(p) &= \slashed{p}\, G_{D^2}(p) + G_{S^2}(p)\,\\
	&= \slashed{p}\, (2G_D G_S) + (G_S^2+p^2G_D^2)\,.
\end{align}
If the original fermion propagator satisfies the KL-representation, \labelcref{eq:Quark-KL} holds as well for the scalar functions $G_{D^2}$ and $G_{S^2}$. 

In both the scalar and fermionic cases, a careful treatment of the product of distributions is required in order to resolve the discontinuities of the squared propagators correctly, for examples see~\cite{Kockler:2025kdt,Pawlowski:2025etp}.

\bibliography{../ref-lib.bib}

\end{document}